\documentclass[conference]{IEEEtran}
\IEEEoverridecommandlockouts
\usepackage{cite}
\usepackage{amsmath,amssymb,amsfonts}
\usepackage{algorithmic}
\usepackage{graphicx}
\usepackage{textcomp}
\usepackage{subfig}
\usepackage{multirow}
\usepackage{float}
\usepackage[utf8]{inputenc}
\usepackage{aeguill}
\usepackage{multirow}
\usepackage{color}
\usepackage{multicol}
\usepackage{enumitem}
\usepackage{lineno}
\def\BibTeX{{\rm B\kern-.05em{\sc i\kern-.025em b}\kern-.08em
    T\kern-.1667em\lower.7ex\hbox{E}\kern-.125emX}}
    
\onecolumn 
    
\pdfoutput=1

\begin{document}

\title{Integrating Proactive Mode Changes in Mixed Criticality Systems\\
}

\author{\IEEEauthorblockN{1\textsuperscript{st} Flavio R. Massaro Jr.  2\textsuperscript{nd}  Edson L. Ursini} 3\textsuperscript{rd}Paulo S. Martins    
\IEEEauthorblockA{\textit{School of Technology} \\
\textit{University of Campinas}\\
Limeira, Brazil \\
frmassaro@gmail.com, ursini@ft.unicamp.br, paulo@ft.unicamp.br}
}

\maketitle

\begin{abstract}

In this work, we propose to integrate prediction algorithms to the scheduling of mode changes under the Earliest-Deadline-First and Fixed-priority scheduling in mixed-criticality real-time systems. The method proactively schedules a mode change in the system based on state variables such as laxity, to the percentage difference in the temporal distance between the completion time of the instance of a task and its respective deadline, by the deadline (D) stipulated for the task, in order to minimize deadline misses.
The simulation model was validated against an analytical model prior to the logical integration of the Kalman-based prediction algorithm. 
Two study cases were presented, one covering earliest-deadline first and
the other the fixed-priority scheduling approach. The results showed the gains in the adoption of the prediction approach for both scheduling paradigms by presenting a significant reduction of the number of missed deadlines for low-criticality tasks.
\end{abstract}


\section{Introduction}
\label{sec:introuction}

Real-time systems are used in critical operations (spaceships, aircraft, vehicles, air traffic control, multi-media communications, etc), where execution time guarantees are required to a task set running in a single or multiple processors. These guarantees can be provided by schedulability analysis. The schedulability analysis computes the worst-case response time (WCRT) for each task. The WCRT is measured by the interval between the release and the completion of each task.  The system is feasible when the WCRT for each task is less than its deadline.

Increasingly, modern systems need to be dynamically adaptable to the operational environment. Thus, changes in the operational environment may require changes in the behavior of the system. These changes can be provided by using multiple operation modes. The modes of operation are represented by a set of tasks optimized to operate in a specific condition. 
Also, the new generations of real-time systems require a design where tasks with different levels of criticality must coexist on the same hardware platform. These features are provided by mixed-criticality real-time systems. In mixed-criticality real-time systems, the tasks are sub-divided in two or more levels of criticality (for example safety critical, mission critical and low-critical). In this context, tasks with a higher level of criticality require hard real-time guarantees, while low criticality tasks could miss her deadlines.

In a real-time system, the correctness of the system behavior depends not only on the logical results of the computations but also on the physical instant at which these results are produced. Real-time systems are classified from a number of viewpoints i.e. hard and soft real time. A missed deadline in hard real-time systems is catastrophic and in soft real-time systems, it can lead to a significant loss or degradation in service. Hence, the predictability of the system behavior is the most important concern in these systems. Predictability is often achieved by either static or dynamic scheduling analysis of real-time tasks.

In many systems, proactive behavior is necessary. Once an overload condition is detected, it is usually
too late to try and remedy the context, leading to missed deadlines.
In particular within the
context of soft real-time systems, where a number of missed deadlines
is allowed. Within the mixed criticality systems framework, 
the abort of low criticality tasks may lead to resources not being fully released back to the system,
leaving it in an inconsistent state. Therefore, the use of proactive scheduling using prediction
algorithms may support a change in criticality that is less amenable to such undesirable effects
(inconsistent states). The assumption is that, by detecting an overload condition ahead of time,
low criticality tasks may be completed and eliminated without compromising the state of the system.
The context of this work is mixed-criticality systems.

In this work, we claim that prediction algorithms may benefit mixed-criticality systems.
in particularly allowing the implementation of a safer elimination of low-criticality tasks.
The immediate abortion of such tasks may be problematic. A potentially more feasible approach
would be to eliminate such tasks within a certain window of anticipation, after they have fully
completed their execution.
with mixed criticality
Therefore, the goal(s) of this work is to introduce:
a) prediction-based control of 
   tasks to prevent missed deadlines;
b) proactive mode-changes and criticality
   level changes.
Although the issue of inconsistent states has been identified by Burns \cite{Burns2014b}, it
has been not addressed in the literature. This work attempts to mitigate such effects
by means of proactive behavior. Thus, it may support the body of work on mixed criticality
systems which heavily depend on the removal of low-criticality tasks.

Specifically, in this work we:
1)  address the use of prediction algorithms for both EDF and FPPS scheduling approaches.
2) validated the simulation model using an analytical model in a first step.
3) use two modes of operation and analyzed the transition between modes.
4)  analyzed the number of missed deadlines with and without prediction in order
to compare the impact of the prediction algorithm.

The research questions addressed in this work are:
1) is it possible to predict the miss of a deadline?
2) which candidate variables should be predicted?
3) anticipating a mode change can reduce the number of missed deadlines?
4) the use of fuzzy in conjunction with the predictor will have better results than using fuzzy in isolation?
5) is it feasible to use prediction to identify the miss of deadlines?

The remainder of this paper is organized as follows:
Section \ref{sec:review}  introduces background and literature review.
Section \ref{sec:comp} addresses the computational model and underlying assumptions.
Section \ref{sec:methods} shows the methods used in this work.
Section \ref{sec:impl} presents the implementation details of the approach.
Section \ref{sec:cases} details the prediction-based use cases.
In Section \ref{sec:disc} we discuss the main results, and
finally, Section \ref{sec:conc} presents the summary and conclusions.

\section{Background and Literature Review}
\label{sec:review}

In this section, we will be presented a structured literature review, based on the concepts previously presented, that shows the state of the art in mode-changes on mixed-criticality real-time systems. 

\subsection{General Mode-Changes in Real-Time Systems}
\label{general_mode_changes}

Mode-changes in real-time systems were introduced by Tindell et al \cite{Tindell1992}. In this work, the authors presented the main concepts that support the mode-change model and it was introduced a simple proposal of scheduling for tasks during a mode-change.
However, to provide the real-time guarantees in design time it was necessary the use of a schedulability analysis that comprises all the context of a mode-change.

Thus, an initial approach of schedulability analysis to FP (fixed-priority) preemptive uniprocessor real-time operating systems was presented by Pedro and Burns \cite{Pedro1998}. Different from analysis of steady-state mode where the calculation of the WCRT for a determined task evaluate just the interference level of tasks with higher priority belonging to a single operation mode, this approach considered the level of interference in both modes (old and new) during a mode-change. However, to assure that tasks meet their deadlines during a mode-change, the new mode tasks are released with an offset on the first execution.

In Real and Crespo \cite{Real2004} the initial approach proposed by Pedro and Burns \cite{Pedro1998} was extended allowing the use of different offsets to the tasks that remain unchanged in both operation modes (old and new). 

Real and Crespo \cite{Real2001a} proposed an algorithm for automatically assigning offsets. The goal of this algorithm was to assigned offsets to tasks prioritizing promptness and warranties of non-infringement of the IPCP (Immediate Priority Ceiling Protocol). The IPCP protocol ensures that lower priority old-mode tasks that share resources do not block resources used by new-mode, higher-priority tasks \cite{Real1999}. Promptness aims to give priority to schedule the new-mode tasks with higher priority at first.
Another approach to assigning offsets was proposed by Martins et al \cite{Martins2016}. This approach aims minimization of mode-change latency and/or offsets using genetic algorithms. The issue of mode-change latency minimization is modeled both as a single and as a multi-objective optimization. The use of genetic algorithms brought flexible ways to use in different situations, enabling the control of a mode-change \cite{Massaro2015}.

The focus of the previously presented work was the use of schedulability analysis on uniprocessor systems. A proposal for schedulability analysis for fixed-priority in mode-change for multi-core systems is shown in Negrean et al \cite{Negrean2011}. This approach considers the dependency between tasks that could be executed in parallel by different processors. Also, it was explored in this work an approach to measure the latency of mode-change highlighting the importance of their minimization to implement a feasible mode-change.

An approach that uses mode change for scheduling on multi-core using EDF (Earliest Deadline First) was presented by Nelis et al \cite{Nelis2011}. The scheduling using EDF prioritizes the execution of tasks with a close deadline. Even though it is an efficient scheduling approach that eliminates the complex step of assigning priority to the task set and reduces the complexity of the system design, it does not override fixed priority when criticality levels are embedded, because some systems do not support the scheduling using traditional EDF \cite{Baruah2008}.

As the use of traditional EDF scheduling on mixed-criticality context is not feasible, another aspect that must be emphasized is the priority assignment that is discussed in detail in Davis et al \cite{Davis2016}. This work presented the most relevant methods and/or algorithms for priority assignment and a new approach was proposed. This new algorithm was compared with the other existing ones through an experiment. The results show that the new proposal had better results than others.

Based on the previously presented context it is possible to see that the use of mode changes can be a feasible way to minimize the design complexity for real-time systems and it reduces the use of resources, as can be observed in Dziurzansk et al \cite{Dziurzanski2016}. In this work, a system using mode change design was implemented in a real hardware platform for a vehicle and shows that the use of operation modes may reduce 75\% of the power consumption of the processors.

\subsection{Mixed-Criticality on Real-Time Systems}  
\label{mixed-Criticality}

Hard real-time constraints are not always necessary for some systems. For example, in a real application, there are tasks with different criticality levels where it is acceptable to miss deadlines on tasks with low criticality. This concept which expected tasks with multiple criticality levels in the same system configuration was introduced by Vestal \cite{Vestal2007}. This work addressed the use of response time analysis for uniprocessors and showed that neither rate monotonic nor deadline monotonic priority assignment were optimal for MCS (Mixed-Criticality Systems), suggesting the use of the algorithm proposed by Audsley \cite{Audsley2001}. In Baruah and Vestal \cite{Baruah2008}, the work presented by Vestal \cite{Vestal2007} was extended to use with sporadic tasks. Also, this approach demonstrated that EDF does not override FP when criticality levels are embedded in some systems.
 
To provide the necessary real-time guarantees for use in real applications, a FP preemptive schedulability analysis for MCS using uniprocessor was proposed in Baruah et al \cite{Baruah2011}. However, this proposal enables the use of only two criticality levels. Thus, a new approach which enables the implementation on multiple criticality levels using multi-core systems was introduced in Pathan \cite{Pathan2012}. Nonetheless, this approach is limited because it does not enable changes during the task period between different criticality levels. In contrast with the approach using fixed priority, other works propose the use of EDF scheduling. However, the employment of a traditional EDF scheduling is not supported in MCS as noted earlier. Thus, it was proposed in Ekberg and Yi \cite{Ekberg2012} that EDF imitates the FP scheme by assigning two relative deadlines to each high criticality task. The first deadline represents the "real" deadline of the task and the second a "virtual" deadline. The virtual deadline is used to increase the probability that high criticality tasks are executed before low criticality tasks. An extension of this work was presented in Ekberg and Yi \cite{Ekberg2014}, where it introduced the use of more than two criticality levels and it enabled changes in the task parameters. Nevertheless, this proposal was designed to be used in uniprocessor systems. A new approach that enables the use of schedulability analysis using EDF in mixed-criticality using multi-core was presented in Baruah et al \cite{Baruah2014}. This work is based on the same concept of virtual deadlines introduced by Ekberg and Yi \cite{Ekberg2012} and it proposes a scheduling algorithm called EDF-VD (for Earliest Deadline First with Virtual Deadlines), whose validation of the theoretical analysis was performed through a simulation. In this proposal, the higher priority tasks have their deadlines reduced (if necessary) during the execution of the low criticality mode.

\subsection{Mode-Changes on Mixed-Criticality Real-Time Systems}  \label{mode_changes}

A discussion on general mode changes and their relation to mixed-criticality systems was presented by Burns (2014). This work listed the main concepts involved and proposed ways to enable the use of mode changes and mixed criticality in the same system, as was shown through an example of application. Also, it concluded that in criticality mode changes, the changes from low to high criticality are closest with degradation mode to accommodate the requirements of a system failure, while the changes from low to high criticality are addressed through functional modes. 
Another relevant proposal which addresses general mode changes and mixed criticality coexisting in the same system was presented by Niz and Phan \cite{Niz2014}. In this approach of partitional multi-core scheduling for multi-modal mixed-criticality real-time systems, it was observed that it is necessary to provide real-time guarantees to tasks with different criticality levels not only in steady state mode but also during a general mode change. Therefore, to accommodate these requirements it proposed a package algorithm and a scheduling algorithm for assigning tasks to processors.

\subsection{Prediction}

Prediction methods are used to prediction an uncertain future state based on previous experiences or knowledge. 
These methods use statistical models that are based on statistical inference principles. In this context, statistical predict inference is one of the approaches that can be used. In this approach, e.g., the previous knowledge of a sample of a population can be used to predict another population related in the same or in different times. Another approach, at where the previous knowledge is provided in a specific or across time is called predicting.
The statistical techniques employed include regression analysis and it their sub-categories such as: linear regression, generalized linear models (logistic regression, Poisson regression, probit regression)  \cite{cox2006}.

In processor scheduling is necessary to predict future states across time, thus, the predicting approach using sequential learning will be the focus of this research. Thus, sequential learning sends to Single Layer Feedforward Network (SLFN) and to the Kalman filter, that will be presented in detail to follow.

\subsubsection{KSL Algorithm}

KSL is a new software reliability growth model (SRGM) based on the Kalman filter with a sub filter and the Laplace trend test present by Ursini et al (2014) \cite{Ursini2014}. The authors applied the model to the Linux operating system kernel as a case study to predict the absolute and relative (per lines of code) number of faults n-steps ahead. The Laplace trend test was applied to detect when the series no longer follow a homogeneous Poisson process, improving the confidence level. An example was provided with a prediction of 13 months ahead on the number of faults with 8\% error. The results (i.e. predictive capability) indicated that the proposed approach outperforms the S-shaped prediction model, Weibull, and Exponentiated Weibull distributions, as well as typical and OS-ELM Neural networks when the series has a short number of observations \cite{Ursini2014}.

Despite the advances in the scheduling theory of real-time systems, the issue
of using prediction algorithms in these
systems have been not addressed.

\section{Computational Model and Assumptions} \label{secModel}
\label{sec:comp}

The simulation model assumes actual execution times, unlike the analytical model, which is based on Worst-Case Response times.
The laxity variable corresponds to the percentage difference in the temporal distance between 
the completion time of the instance of a task and its respective deadline, by the deadline 
(D) stipulated for the task, will be used as the basis for the object of prediction, i.e. 
$Lax = D_i-R/D_i$, where $R$ is the simulated response time for a task $\tau_i$
and $D_i$ is its deadline.

\begin{figure}[h]
    \centering
    \hspace{-2.5cm}   
 \includegraphics[width=10cm]{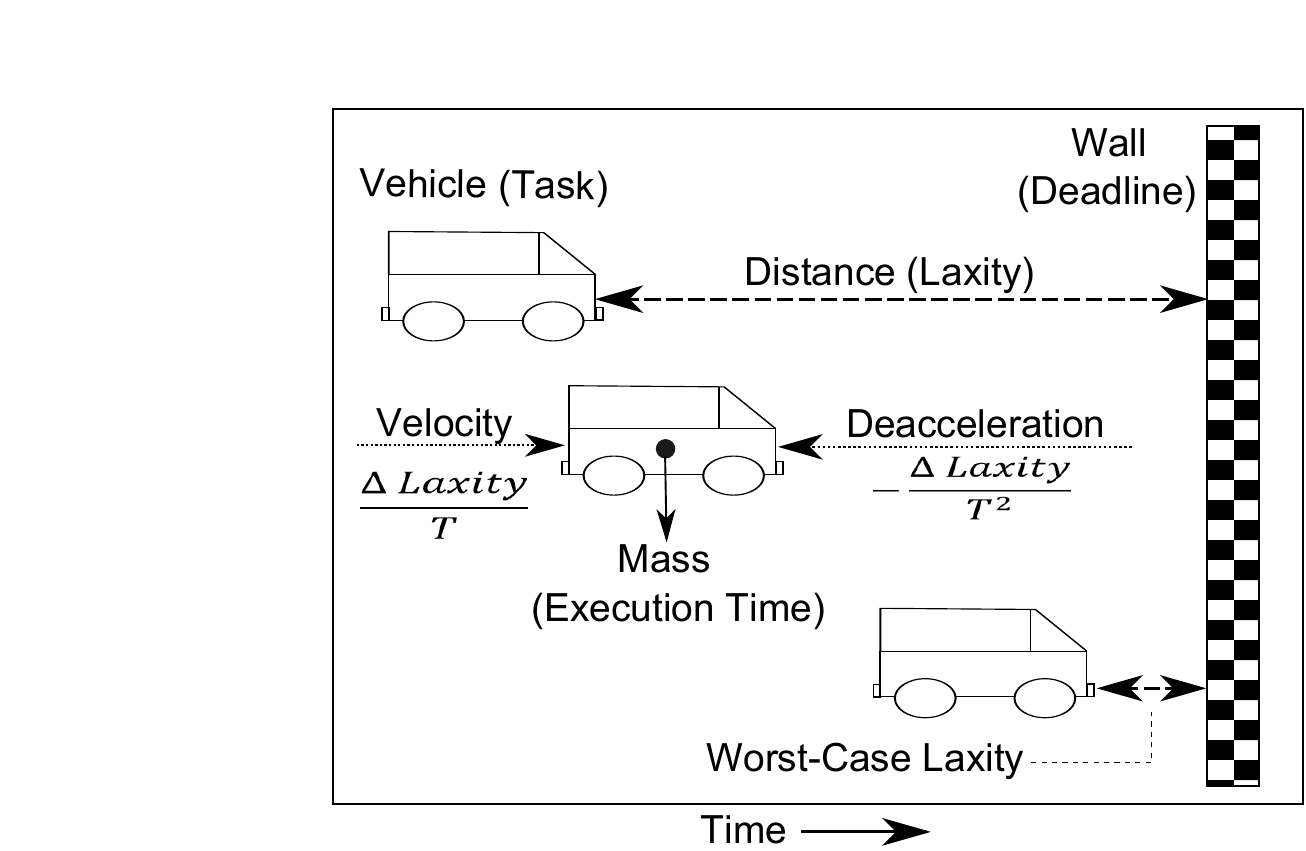}    
    \caption{\label{fig:Laxity} Model analogy: variable for predicting miss of deadline}
\end{figure}

Fig. \ref{fig:Laxity} represents an analogy between the miss of deadline of a task with the collision of a moving vehicle with another object. In this analogy, it can be observed that the more the vehicle approaches the object, the greater the probability of collision. Besides the distance, another factor that influences the moment of the collision is the current vehicle speed. The correlation between proximity and speed indicates whether the vehicle has a tendency to collide or not. Based on this analogy, it was stipulated that the less laxity of tasks completed in a given time window will be used as the object of prediction. However, it should be noted that the window interval is a parameter that can vary according to the system configuration. Completing the analogy with the vehicle in motion, velocity is another important variable that must be used in the prediction of the miss of deadline. To calculate the velocity value, the angular coefficient of the last two observations of the lowest laxity is calculated. 

Therefore, the lower laxity variables in a given time window and the slope of the last two observations will serve as the basis for determining if the system is prone to imminent miss of deadline.

Based on the above, this work is expected to predict the imminent miss of a deadline so that preventive measures can be taken in a timely manner in order to minimize the impact caused by this miss.


The computational model is composed of the following elements:
1) task generator module,
2) CPU scheduling (EDF,FP) module,
3) analyzer module,
4) system predictor (pro-activity) module,
5) mode-manager module,
6) task dispatcher module.

\begin{figure}[h]
     \centering
  \includegraphics[angle=0,scale=0.4]{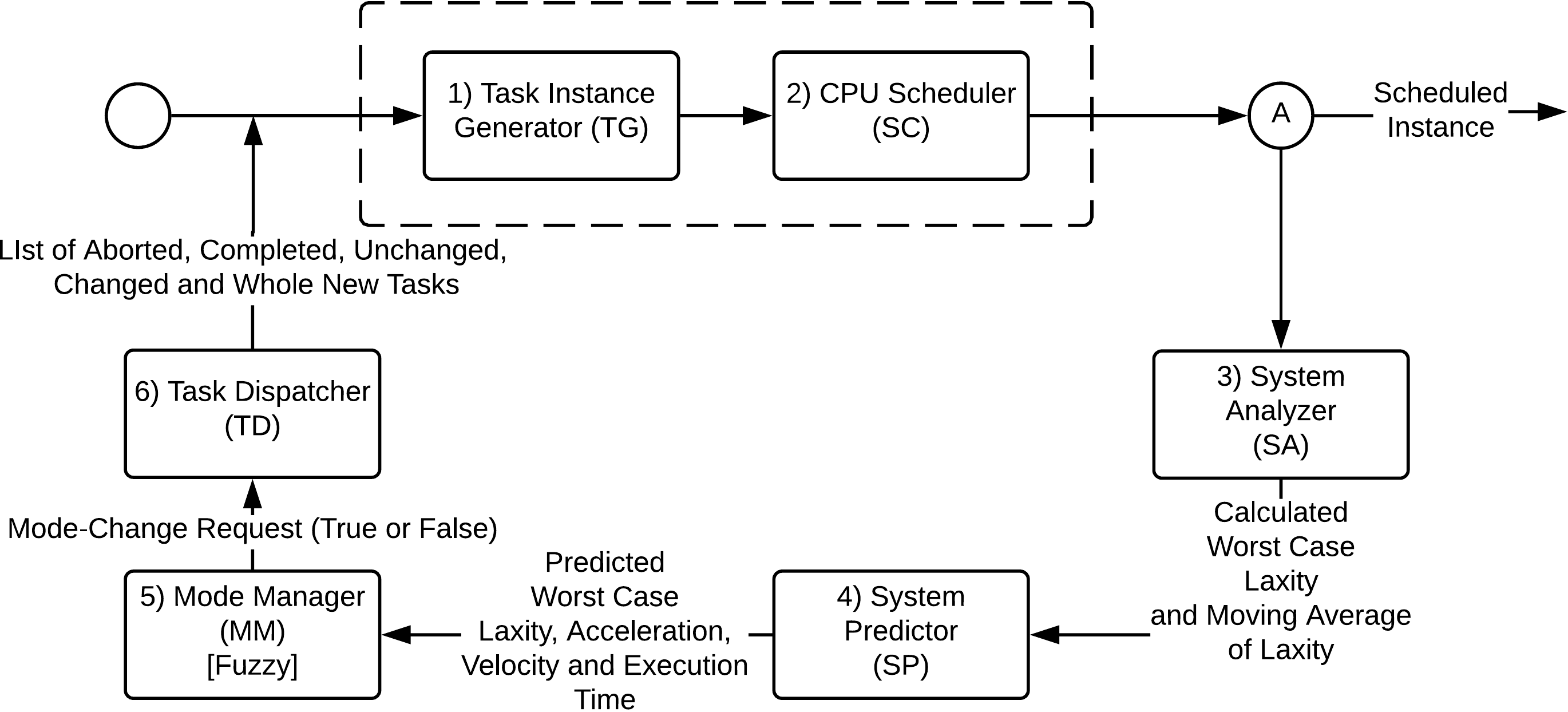}    
\caption{\label{fig:proModel} Top-level view: proactive model}
\end{figure}

Fig. \ref{fig:proModel} shows all the major components of the simulation model. The simulation model is sub-divided into six blocks: 

\begin{enumerate}[leftmargin=*]

\item \textit{tasks generator}: this block generates the tasks according to the periodicity required and stores the owner mode and the start time for each task released;
\item \textit{CPU scheduling}: this block puts the released tasks on a processing queue when an arriving task has a priority higher than that of the task in execution, is performed the preemption of the executing task; 
\item \textit{analyzer}: this block account the variables used for system predictor such as worst case laxity, velocity, etc;
\item \textit{system predictor}: this block is responsible for predicting the events that trigger a mode change;
\item  \textit{mode manager}: this block is used to manage the mode change process, whose start decision is obtained from the result of a fuzzy function that uses as input parameters the worst case laxity and the coefficient of the last two measurements;
\item \textit{task dispatcher}: computes and sends the list of aborted, completed, unchanged, changed and whole new tasks to the task generator.

\end{enumerate}

\begin{figure}[h]
    \centering
    \includegraphics[angle=0,scale=0.4]{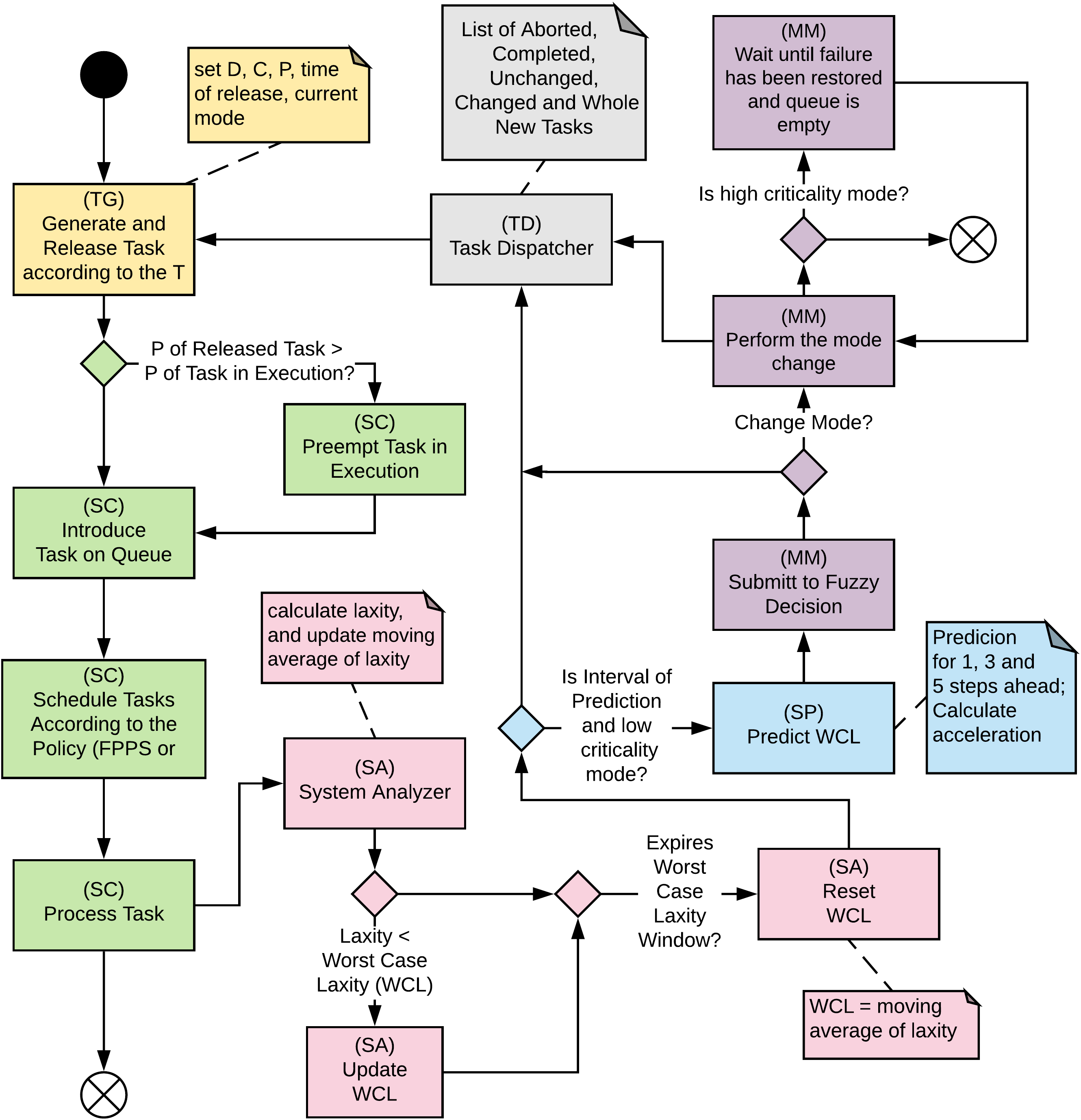}    
    \caption{\label{fig:SimDiag} Expanded View of the Proactive Model}
\end{figure}

Fig. \ref{fig:SimDiag} presents an expanded view of the proactive model. In this figure, each color represents a module previously  described.

\subsection{Mode-Change Tasking Model}
In this work was addressed a set of periodic or sporadic tasks  $ \tau = \{ \tau_1,\tau_2, \ldots \tau_i, \ldots  \tau_p \}$ per mode. Each task $\tau_i$ is 
composed by the record  $S_i \ = \ \{T_i,D_i,C_i,P_i\}$, where: 1) $T_i$ and $D_i$ are respectively the period of task $\tau_i$ 
(or, if a sporadic task, the minimum inter-arrival time between successive tasks of the stream $i$) and the deadline; 2) $C_i$ is the worst-case execution time (WCET) of the task $\tau_i$. This value is deemed to contain the overheads due to context switching. Moreover, the values of $C_i$, $D_i$ and $T_i$ are such that $C_i$ < $D_i$ $\leq$ $T_i$ . 
Also, it was not considered the restriction that$D_i$ $\leq$ $T_i$ ; 3) $P_i$ represents the priority of task $\tau_i$ , assigned according  to the Deadline Monotonic Scheduling algorithm.

The mode change model is based on the same  assumptions of Real and Crespo \cite{Real2001} and Pedro and Burns \cite{Pedro1998}, i.e., 
1) tasks are assigned fixed priorities by the Deadline Monotonic Scheduling algorithm (DMS) for Fixed Priority Policy, for Earliest Deadline First it is dynamically assigned  
2) tasks are executed in a uniprocessor system;
3) tasks are not permitted to voluntarily suspend themselves during an invocation;	
4) there are fixed task sets before and after the mode change, and
5) tasks are scheduled with time offsets during the mode change only. This time phasing between tasks may or may not hold after the mode change. 

\subsection{Schedulability analysis for Fixed-Priority} 

In this section we will provide a schedulability analysis that has been used to calculate the worst-case latency of a task across a mode change that will be used as validation of the simulation model. For more details on the basic fundamentals of real-time schedulability analysis, the reader is referred to Burns et. al  \cite{Burns2009}.

 There are two types of task offsets $O_{i(N)}$:  $Y$ is measured from the start of a mode change (MCR) and
$Z$ is measured from the end of the period of the preceding old-mode task \cite{Real2001}. 
The  window $x$  is the phasing between the MCR and the activation of task $\tau_i$. 

\subsubsection{Analysis for Old-mode Tasks}
The interference level of old mode tasks  is given in accordance with its classification of the types of tasks:
1) Interference from higher priority old mode completed tasks $I_{hp(i)_O}$;
2) Interference from higher priority aborted tasks $I_{hp(i)_A}$;
3) Interference from higher priority new mode tasks $I_{hp(i)_N}$, 
and 4) Interference from unchanged tasks  $hp(i)_U$.
By combining the analysis of interference of each task type we obtain the analysis model of the old mode, given by the following  recurrence equation:

\begin{equation}\label{eq:equacao_OLD_WN+1}
\begin{gathered}
w_i^{n+1}=C_i+B_i+\sum _{\forall j\in hp(i)_O}\left \lceil \frac{x}{T_j} \right \rceil C_j+\\
\sum _{\forall j\in hp(i)_A} \left ( \left \lfloor \frac{x}{T_j} \right \rfloor C_j + min \left ( x - \left \lfloor \frac{x}{T_j} \right \rfloor T_j,C_j \right )\right  ) + \\
\sum_{\forall j\in hp(i)_N}\left \lceil \frac{w_i^n-x-Y_j}{T_j} \right \rceil_0 C_j+\\
\sum_{\forall  j\in hp(i)_U}\left \lceil \frac{x}{T_j} \right \rceil C_j + \left \lceil \frac{w_i^n- \lceil x/T_j \rceil T_j-Z_j}{T_j} \right \rceil_0 C_j \\
\end{gathered}
\end{equation}

The notation $\lceil z \rceil_0$ denotes a modified ceiling function that returns zero if $Z<0$.
The initial value of $w_i$ is set to zero. It can be shown that  $w_i^{n+1}>w_i^n$, and hence the equation is 
guaranteed either to converge (i.e. $w_i^{n+1}=w_i^n$) or to exceed some threshold, such as $D_i$. However, 
the worst case for the response task $R_i$, which must then be compared with the respective deadline is given by  $R_i$ \ = \ $w_i$ \ + \   $C_i$.

\subsubsection{Analysis for New-mode Tasks}
Because new mode tasks suffer from interference from other old and new higher priority tasks, we need to guarantee
heir schedulability during the mode change. If, however, a new task
$\tau_i$ has an offset such that its first release occurs after all higher priority old mode tasks have completed, its schedulability is 
guaranteed by steady-state analysis and we do not need to apply the following analysis to obtain its WCRT.  The interference suffered
from a new mode task is analyzed as follows:
1) Interference from higher priority old mode tasks ${hp(i)_O}$;
2) Interference of the tasks belonging only to the new mode $I_{hp(i)_N}$, and
3) Interference of unchanged tasks  $hp(i)_U$.
The worst case response time of a new task $i$ across a mode change is therefore given by:

\begin{equation}\label{eq:equacao_NEW_WIN}
\begin{gathered}
 w_i^{n+1}=B_i+\sum _{\forall j\in hp(i)_O}C_j+
\sum _{\forall j\in hp(i)_N}\left \lceil \frac{w_i^n-Y_j}{T_j} \right \rceil_0 C_j+\\
\sum _{\forall j\in hp(i)_U}\left( C_j + \left \lceil \frac{w_i^n-T_j-Z_j}{T_j} \right \rceil_0 C_j \right ) \\ \\
\end{gathered}
\end{equation}

The initial value of $w_i$ is set to zero. It can be shown that  $w_i^{n+1}>w_i^n$, and hence the equation 
is guaranteed either to converge (i.e. $w_i^{n+1}=w_i^n$) or to exceed some threshold, such as $D_i$. 
However, the worst case for the response task $Ri$, which must then be compared with the respective deadline is given by  $R_i$ \ = \ $w_i$ \ - \   $Y_i$.

\section{Methods}
\label{sec:methods}
%
In this section will be described the methods used in this work. The methods were divided into four steps:

\begin{enumerate}[leftmargin=*]

\item \textit{Build the simulation model:} Initially, the simulation model was built using the assumptions and the computational model described in the previous section.

\item  \textit{Validate the simulation model:} To validate the simulation, a case study was performed to in single-mode and multimode. The results obtained, regarding the response times of each of the tasks, were compared with the analytical model (schedulability analysis equations).

\item  \textit{Predict mode-changes:} After the evaluation of the simulation model, two studies were elaborated, one using EDF scheduling policy and another using FP. These case studies were used to evaluate the effectiveness of the proposed approach, ie, prediction of mode-changes in mixed-criticality real-time system. In each case study, single mode, reactive multimode and multi-mode proactive simulations were performed. In addition, for the purpose of comparative analysis, simulations were performed using the prediction module, but not executing mode changes.

\item \textit{Analyze the results:} The results of the simulations carried out in each case study were tabulated and analyzed to support the discussions and conclusions of this work.

\end{enumerate}

\section{Implementation}
\label{sec:impl}


In this section was implemented the model showed in 
Figures \ref{fig:proModel} and \ref{fig:SimDiag}, 
through discrete event simulation.

\subsection{Mode-Change Simulation and Validation}  
\label{simulation_model}

The focus of this section is to propose a practical solution for the use of a simulation model for mode changes in preemptively fixed-priority real-time systems, which involves the modeling of the approach and its validation.

The simulation model used in this section was divided into two parts: 1) steady-state and 2) mode change. In steady-state is the system was simulated using just a single mode. The mode change model extends the steady-state mode to allow the simulation of a multi-mode system.

The simulation was performed on a set of six generic tasks  (Table \ref{tab:SixTasks}).
The goal of this case is twofold
1)  to demonstrate the flexibility of the approach and 
2) to compare the simulation results with the results obtained using an analytic model. 
Therefore, we used the same task set used by Real and Crespo \cite{Real2001}. 
 Table \ref{tab:SixTasks} presents two modes of operation $M1$ and $M$2; 
 each mode of operation is made up of six tasks, i.e.  $\tau_1$ through $\tau_6$.  
 The CPU utilization rate is $73.1\%$ in $M1$ and $66.35\%$ in $M2$. 
Note: The column "TEST" shows that the task is feasible in steady-state mode.

\begin{table}[htp]
\centering
\setlength{\tabcolsep}{0.5pt}

\caption{Case study - Task set}	

\begin{tabular}{|c|c|c|c|c|c|c|c|c|c|c|c|c|c|c|c|c|c|} \cline{1-8} \cline{10-17}
\multicolumn{8}{|c|}{\textbf{Mode M1}} & & \multicolumn{8}{c|}{\textbf{Mode M2}} \\ \cline{1-8}	\cline{10-17}
	
{\textbf{Tasks}} & \textbf{P} & \textbf{C} & \textbf{B} & \textbf{T=D} & \textbf{O} & \textbf{R} & \textbf{TEST} & & {\textbf{Tasks}} & \textbf{P} & \textbf{C} & \textbf{B} & \textbf{T=D} & \textbf{O} & \textbf{R} & \textbf{TEST} \\ \cline{1-8} \cline{10-17}
$\tau_{1 (O)}$ & 1 & 10 & 0  & 100 & 0 & 10  & OK 
        & & $\tau_{1 (U)}$ & 1 & 10 & 0  & 100 & 0 & 10  & OK \\ \cline{1-8} \cline{10-17}
$\tau_{2 (W)}$ & \multicolumn{7}{|c|}{not active in this mode} 
        & & $\tau_{2 (W)}$ & 2 & 20 & 0 & 120 & 195 & 30  & OK \\ \cline{1-8} \cline{10-17}
$\tau_{3 (O)}$ & 2 & 30 & 0 & 200 & 85 & 40  & OK 
        & & $\tau_{3 (C)}$ & 3 & 30 & 0 & 270 & 0 & 60  & OK \\ \cline{1-8} \cline{10-17}
$\tau_{4 (O)}$ & 3 & 40 & 0 & 280 & 0 & 80 & OK 
        & & $\tau_{4 (U)}$ & 4 & 40 & 0 & 280 & 70 & 100 & OK \\ \cline{1-8} \cline{10-17}
$\tau_{5 (O)}$ & 4 & 50 & 0 & 300 & 0 & 140 & OK 
        & & $\tau_{5 (C)}$ & 5 & 50 & 0 & 350 & 0 & 180 & OK \\ \cline{1-8} \cline{10-17}
$\tau_{6 (O)}$ & 5 & 60 & 0  & 350 & 100 & 200 & OK 
        & & $\tau_{6 (O)}$ & \multicolumn{7}{|c|}{not active in this mode}\\ \hline
\end{tabular}

\label{tab:SixTasks}
\end{table}

The simulations were performed on steady-state and mode change models. In steady-state, the simulation was performed during  $4 \times 10^2$ time units and in mode change during $4 \times 10^6$ time units and the interval between mode changes is given by a random exponential distribution with a value equal to $2 \times 10^3$ time units.

First, the simulation was built for a steady-state model based on a generic model. By definition, the system is in steady state when it executes a fixed task set (without mode changes) \cite{Massaro2015}. Thus, this model uses a single-mode which simplifies its implementation.

\begin{figure}[htp]
  \centering   
  \subfloat[fig:WCRT_RSS_M1][Mode M1]{\includegraphics[width=6cm]{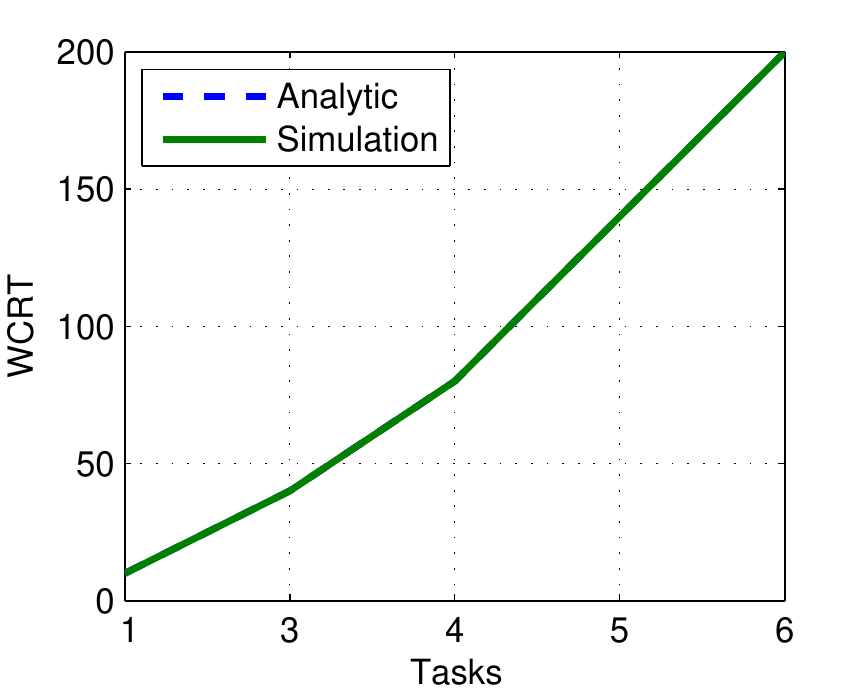}}  
  \hfill
  \subfloat[fig:WCRT_RSS_M2][Mode M2]{\includegraphics[width=6cm]{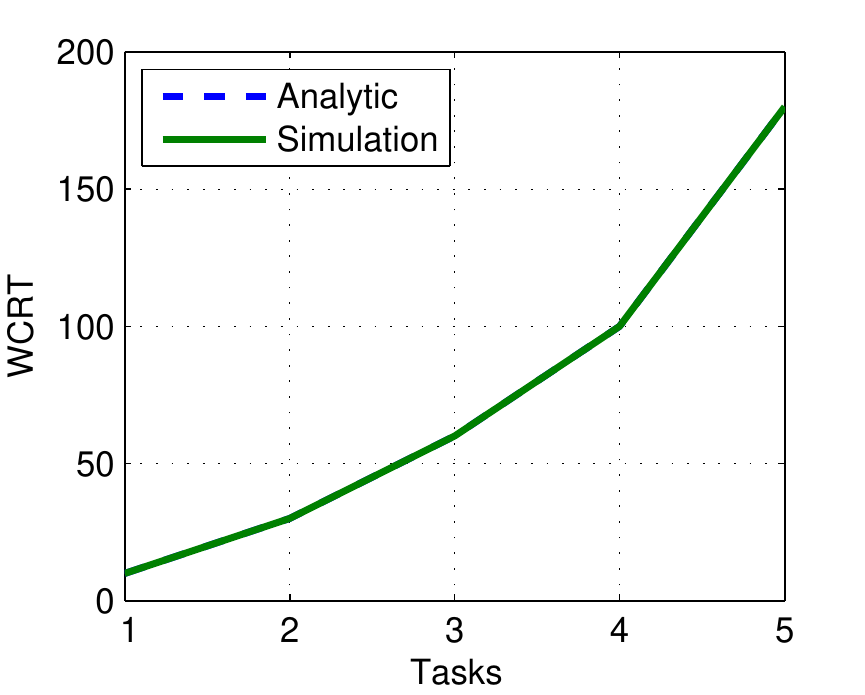}} 
  \caption{\label{fig:WCRT_RSS} WCRT in steady-state mode}
\end{figure}

Fig. \ref{fig:WCRT_RSS} shows a comparison between analytic and simulation results for the worst-case response time in steady-state mode. As we can see in both figures below, the simulation results for both modes were the same as for the analytic model. These results validate the simulation for the steady-state model.
 
\begin{figure}[h]
  \centering    
  \subfloat[fig:WCRT_M1M2_A][Old mode tasks]{\includegraphics[width=6cm]{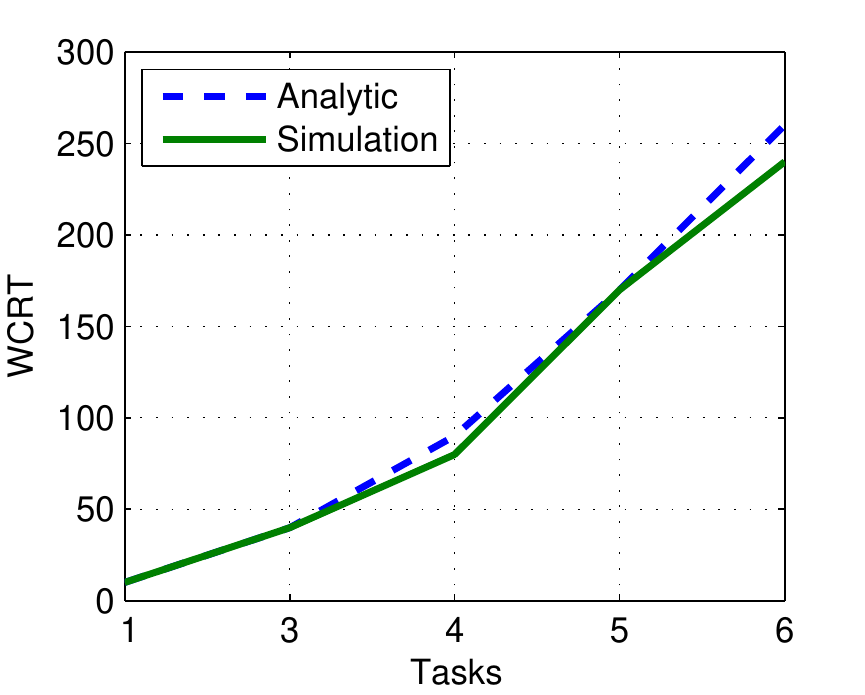}}  
  \hfill
  \subfloat[fig:WCRT_M1M2_B][New mode tasks]{\includegraphics[width=6cm]{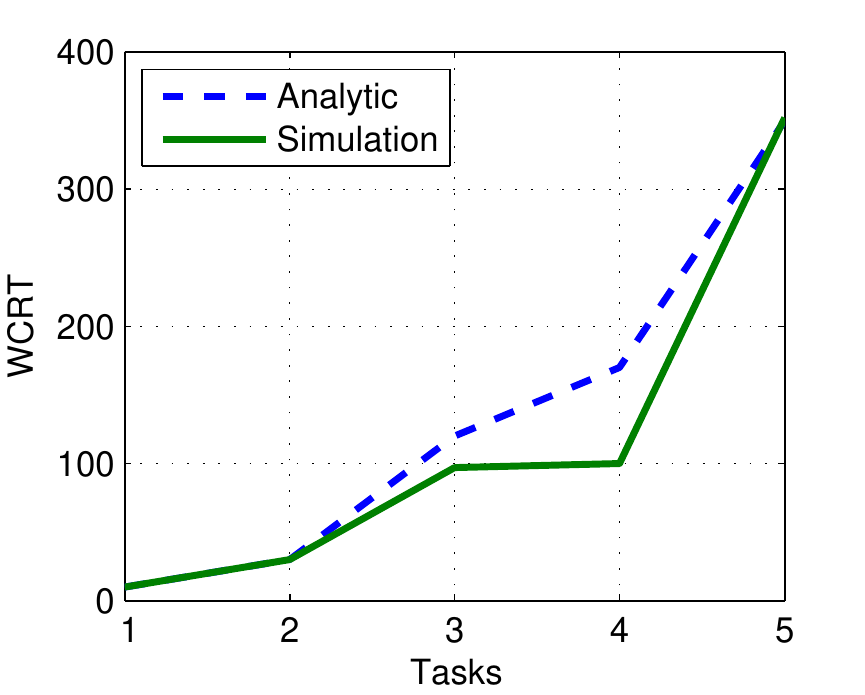}}   
  \caption{\label{fig:WCRT_M1M2} WCRT in mode-change from M1 to M2}
\end{figure}

\begin{figure}[h]
  \centering   
  \subfloat[fig:WCRT_M2M1_A][Old mode tasks]{\includegraphics[width=6cm]{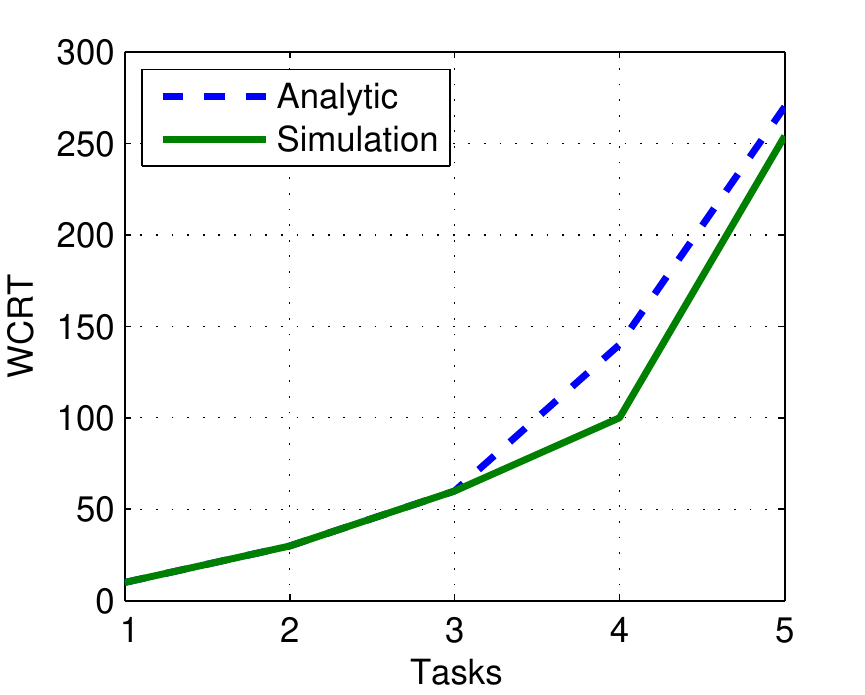}}  
  \hfill
  \subfloat[fig:WCRT_M2M1_B][New mode tasks]{\includegraphics[width=6cm]{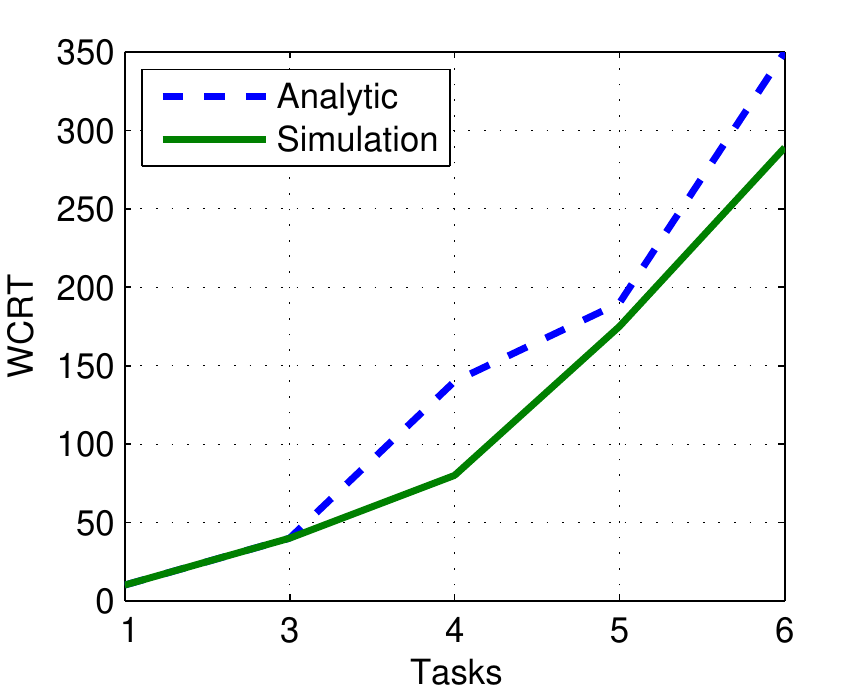}} 
  \caption{\label{fig:WCRT_M2M1} WCRT in mode-change from M2 to M1}  
\end{figure}

Fig. \ref{fig:WCRT_M1M2} shows a comparison between analytic and simulation results of worst-case response times for mode changes from mode M1 to mode M2 and Fig. \ref{fig:WCRT_M2M1} from mode M2 to mode M1. As we can see in both figures, the simulation results for all old-mode tasks were close to the analytic model. However, it is possible to see a large difference in new-mode task  $\tau_4$ in both modes. Also, we may emphasize that in new-mode tasks 5 the worst-case response time observed in the simulation from M1 to M2 was slightly longer than that of the analytic model with a value of 0.57\% higher than their deadline.
 
Besides the worst-case response time in mode change, another relevant indicator is the mode latency. The worst-case latency observed in the simulation was $399$ms from mode M1 to mode M2 and $389$ms from mode M2 to mode M1. Likewise, the worst-case latency calculated using the analytic method was $420$ms from mode M1 to mode M2 and $450$ms from mode M2 to mode M1. The results of worst-case latency in mode change were close in a comparison between simulation and analytic methods with differences in the order of $5$\% and $13.6$\%, respectively.

As analytic models are very limited for the general characteristics of real-time systems, in this section, we have proposed a simulation model to real-time systems in a mode change. It validated the simulation in relation to the analytic model. We argue that the results seem relatively appropriate for general use in simulation models. Therefore, the proposed model can be a flexible approach for work with complex models in mode change, e.g. it could be extended to enable a quick analysis of the behavior of a system configuration in the presence of any faults. 

\section{Predicting Mode-Changes in Mixed-Criticality Systems}
\label{sec:cases}

In this section, will be presented mechanisms that make it possible to anticipate mode-changes in a mixed-criticality real-time system, using prediction.
This section is subdivided into two case studies described as follows: 1) predicting mode-changes in EDF and 2) predicting mode-changes in FP.

 \subsection{Case Study 1 - Predicting Mode-Changes in EDF}
\label{case1}

One of the assumptions of mixed-criticality systems is that when a change in the criticality level occurs, the least critical tasks that are in the queue or in processing are aborted immediately or that they may have execution within their compromised commit if necessary, to remain active in the system \cite{Burns2014b}. This assumption may in certain cases compromise the execution of the system when there are tasks with low criticality, whose execution is necessary to release certain resources of the system. Thus, the prediction of a mode-change may be favorable because, with the detection of the imminent arrival of a criticality level change, it would be possible to block new instances of the low criticality tasks, allowing existing instances to complete their execution. Thus, when the request for change of the criticality level occurs, the system will have sufficient resources to respond promptly to this change, reducing even the mode-change latency. If the prediction is not carried out and the system does not perform the mode-change, it is necessary to remove the lock of a release of new instances of the tasks with a low level of criticality aand to interrupt the release of only tasks of the high level of criticality.

\subsubsection{Assumptions}

Using simulation, a test was performed to verify the effectiveness of the deadline prediction process. In this case study, the following assumptions were adopted: 1) the system used EDF scheduling policy, 2) the predictor used was KSLX, 3) the variable to be monitored was the worst case laxity, which corresponds to the percentage of the shortest distance for the deadline of the tasks performed in relation to their respective deadline during a measurement window, which in this case study corresponds to $ 20 $ MDC (T), 4) the prediction was performed every time interval, determined by the maximum common divisor (MDC) of the periods (T) of the tasks, 5) fuzzy logic was used to determine the imminent miss of deadline, 6) processor failures were randomly generated. To measure the accuracy level of the prediction, the root mean square error (RMSE) was used as the basis between the predicted value and the real value, that is, the square root of the mean square error and 7) a offset was assigned to the tasks, belonging exclusively to the low criticality level. The assigned offset corresponds to the time difference between the start of the mode-change and the next multiple of the period (T) of the task.

\subsubsection{Configuration}

Table \ref{tab:Case2TaskSet} displays the set of tasks that were used in the simulation. The columns of the table correspond to the task number, criticality level (low or high), task priority at low criticality level, task priority at high criticality level, task criticality low at task level, time of execution of the task in high criticality level, the period/deadline of the tasks, which in this case will be the same. The CPU utilization rate for high criticality level is $73.10\% $ for criticality level down $41.67\% $

To generate the processing failures, an exponential distribution with an average of $10$ and a normal distribution with a mean of $12$ and a standard deviation of $7$ were used as the arrival rate. As a premise, the generation of failures was released after the system warm-up period, set to $2000$ units of time. As the high criticality level is used for system recovery, it was adopted that after the transition from the low criticality level to the high level was completed, the generation of errors was suspended and resumed after the transition from the high criticality level to the low criticality level.

\begin{table}[h]

\begin{minipage}{0.45\linewidth}
\centering
\setlength{\tabcolsep}{1pt}
\caption{Task Set}	
\begin{tabular}{|c|c|c|c|c|c|c|} 
\hline	
\textbf{Task} & \textbf{Level} & \textbf{$P$} & \textbf{$P$}  & \textbf{$C$} & \textbf{$C$} & \textbf{$T=D$}  \\
 &  & \textbf{$(LO)$} & \textbf{$(HI)$}  & \textbf{$(LO)$} & \textbf{$(HI)$} &   \\
\hline	
$\tau_{1}$ & High & 1 & 1 & 10 & 10 & 100 \\ \hline	 
$\tau_{2}$ & High & 2 & 2 & 30 & 30 & 200 \\ \hline	
$\tau_{3}$ & Low & 3 & - & 40 & - & 280 \\ \hline	
$\tau_{4}$ & High & 4 & 3 & 50 & 50 & 300 \\ \hline	
$\tau_{5}$ & Low & 5 & - & 60 & - & 350 \\ \hline
\end{tabular}
\label{tab:Case2TaskSet}
\end{minipage}
\hspace{0.4cm}
\begin{minipage}{0.45\linewidth}
\centering
\setlength{\tabcolsep}{1pt}
\caption{Fuzzy Rules}	
\begin{tabular}{|c|c|c|c|} 
\hline
\multirow{2}{*}{\textbf{Acceleration}} &  \multicolumn{3}{|c|}{\textbf{Prediction}} \\ \cline{2-4} 
            & \textbf{\textit{Ultra}} & \textbf{\textit{Short}} & \textbf{\textit{Normal}} \\
\hline	
\textbf{Fast} & \textit{High}  & \textit{High} & \textit{Low} \\ \hline
\textbf{\textit{Medium}} & \textit{High}  & \textit{Low} & \textit{Low} \\ \hline
\textbf{\textit{Slow}} & \textit{High} & \textit{Low} & \textit{Low} \\ \hline
\textbf{\textit{Negative}} & \textit{High} & \textit{Low} & \textit{Low} \\ \hline
\end{tabular}
%
\label{tab:C2FuzzyRules}
\end{minipage}
\end{table}

The fuzzy logic adopted to decide if the system should be submitted to the mode-change has 2 input variables: 1) acceleration: high, medium, low and negative, which corresponds to the angular coefficient of the line obtained from the last two observations of the variable used as the object of the prediction, 2) distance predicted in percentage: short, short and norm. The fuzzy rules implemented by the KSLX algorithm determine if the prediction used will be 1, 3 or 5 steps forward.

\begin{figure}[h]
  \centering
    
\subfloat[fig:C2FuzzyAcceleration][Acceleration]{\includegraphics[width=6cm]{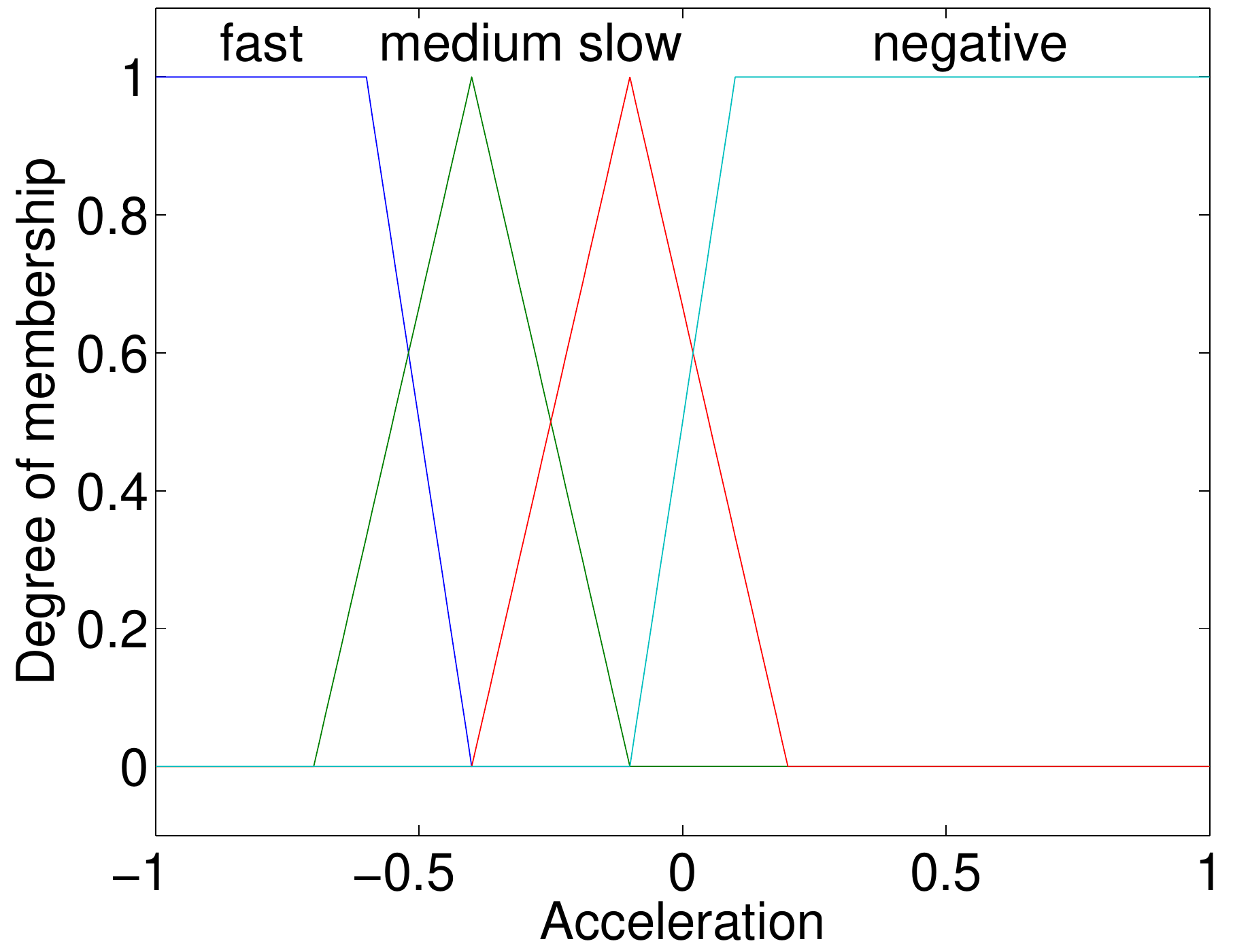}} 
\hfill
\subfloat[fig:C2FuzzyDistPred][Predict Laxity]{\includegraphics[width=6cm]{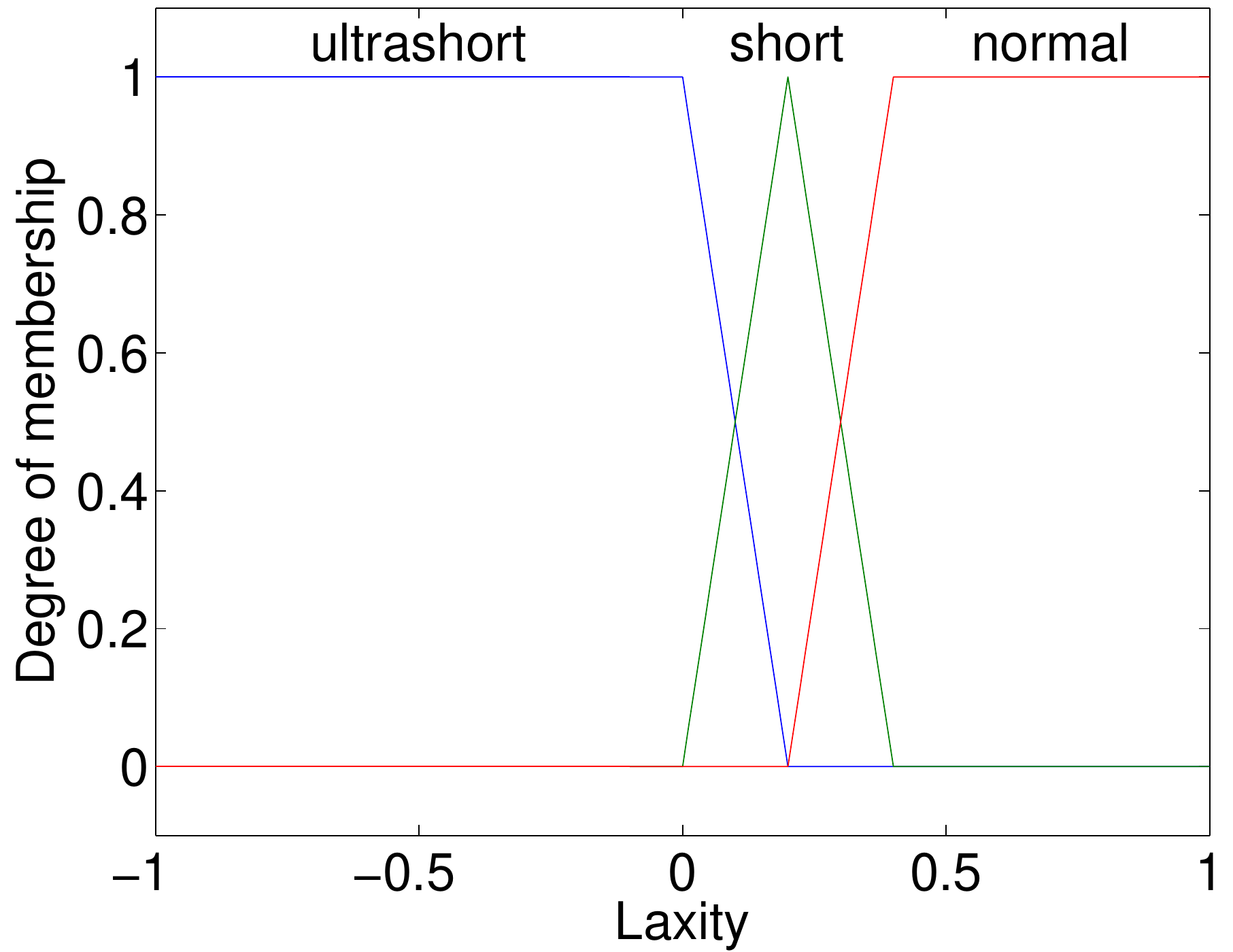}} 

\subfloat[fig:C2FuzzySaida][Fuzzy Output Variable]{\includegraphics[width=6cm]{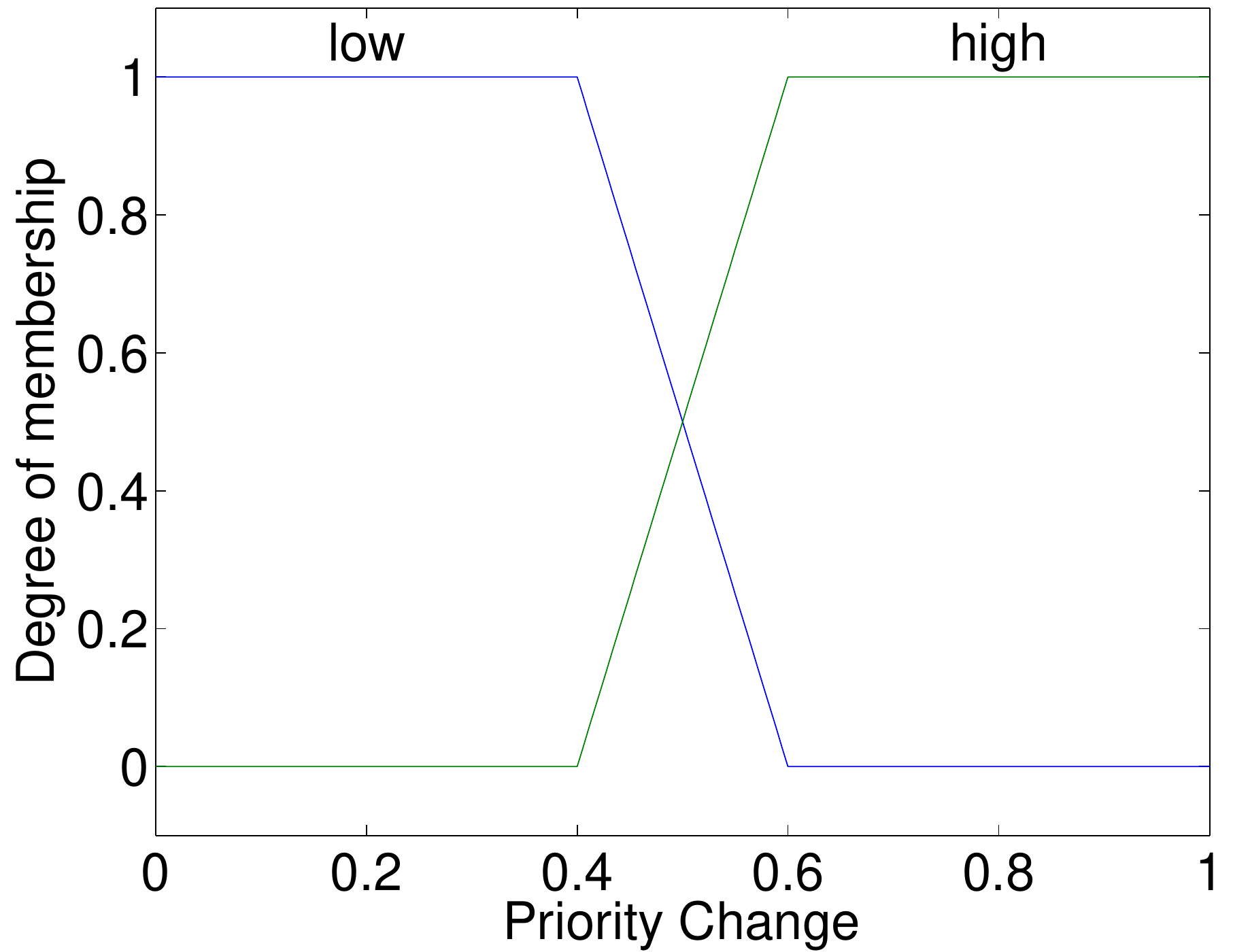}}

  \caption{\label{fig:C2fuzzyInputVariables}Fuzzy Input and Output Variables}
  
\end{figure}

Figs. \ref{fig:C2fuzzyInputVariables}a, \ref{fig:C2fuzzyInputVariables}b present the functions used to subdivide the input groups to the variables related to the fuzzy logic. For the acceleration groups shown in Fig. \ref{fig:C2fuzzyInputVariables} a two trapezoidal functions with respective parameters were used: [-1 -1 -0.6 -0.4] and [-0.1 0.1 1 1] and two triangular functions with the following parameters: [-0.7 -0.4 -0.1] and [-0.4 -0.1 0.2]. Also, as can be seen in Fig. \ref{fig:C2fuzzyInputVariables} b, for sub-division of the groups corresponding to the predicted distance two trapezoidal functions were used with the following parameters: [-1 -1 0 0.2] and [0.2 0.4 1 1] and a triangular function with the following parameters: [0 0.2 0.4].

Fig. \ref{fig:C2fuzzyInputVariables}c demonstrates the output functions that determine the risk degree of miss a deadline. The output values were subdivided into two groups: low and high, represented by trapezoidal functions with the following parameters: [0 0 0.4 0.6] and [0.4 0.6 1 1]. Thus, the result of the defuzzification was used to determine if the system is minded to the imminent miss of deadline.

Table \ref{tab:C2FuzzyRules} presents the 12 fuzzy rules created to determine the degree of miss of deadline for the simulation using the KSLX predictor. The rules determined here, as well as the parameters of the fuzzy functions, should be analyzed and adjusted depending on the application and/or system configuration.


In order to allow an analysis of the advantages obtained by using the predictor, the same system was submitted to another similar simulation process, using only fuzzy logic to identify the need to change the level of criticality.

After the configuration of the FUZZY rules, the simulation processes were executed (with and without the use of the predictor and with a change of the criticality level in a reactive way), with a duration of execution corresponding to 100 hyper-periods with 10 repetitions . Where each  hyper-period, which is determined by the least common multiple (MMC) of the task set periods previously presented in the Table \ref{tab:Case2TaskSet}, is equivalent to 4200 units of time. Therefore, the total time of simulation was  $4200 \times 10^{3}$  units of time. 
 
\subsubsection{Results}

Table \ref{tab:C3Results} presents the results chosen from the simulation process. In this table, the lines represent the measured variables and the columns present the results of each of the simulation process types executed, namely: 1) reactive mode-change, 2) proactive mode-change, using fuzzy in isolation and 3) change of pro-active critique, combining fuzzy with the predictor KSLX. For each type of simulation process two simulations were performed, the first without changing the criticality level and the second with change.

\begin{table}[t]
\centering
\small
\setlength{\tabcolsep}{0.5pt}
\caption{Case Study 1 - Results}
\label{tab:C3Results}

\begin{tabular}{|c|c|c|c|c|c|c|}
\hline
\multirow{3}{*}{\textbf{Variables}}                 & \multicolumn{6}{c|}{\textbf{Results}}                                                                                   \\ \cline{2-7} 
                                                    & \multicolumn{2}{c|}{\textbf{Reactive}} & \multicolumn{2}{c|}{\textbf{Fuzzy}} & \multicolumn{2}{c|}{\textbf{Fuzzy+KSLX}} \\ \cline{2-7} 
                                                    & \textbf{Mono}      & \textbf{Multi}      & \textbf{Mono}     & \textbf{Multi}    & \textbf{Mono}       & \textbf{Multi}       \\ \hline
\textbf{CPU utilization rate}                       & $99.19\%$           & $95.71\%$         & $99.19\%$          & $93.72\%$      & $99.19\%$            & $70.96\%$         \\ \hline
\textbf{CPU Busy}                                   & $73.10\%$           & $70.81\%$        & $73.10\%$          & $70.01\%$      & $73.10\%$            & $58.87\%$         \\ \hline
\textbf{CPU Downtime}                               & $26.09\%$           & $24.90\%$        & $26.09\%$          & $23.71\%$      & $26.09\%$            & $12.09\%$         \\ \hline
\textbf{Number of Completed Tasks}                  & $104,050$           & $102,007$        & $104,050$          & $100,857$      & $104,050$            & $92.064$          \\ \hline
\textbf{Number of misses deadlines}                 & $7.030$             & $753$            & $7.030$            & $474$          & $7.030$              & $227$             \\ \hline
\textbf{Percentage  of misses deadlines}            & $6.75\%$            & $0.74\%$         & $6.75\%$           & $0.47\%$       & $6.75\%$             & $0.24\%$          \\ \hline
\textbf{MSE  - Prediction 1 step ahead}             & N/A                 & N/A              & N/A                & N/A            & $0.0070$             & $0.0079$          \\ \hline
\textbf{MSE  - Prediction 3 steps ahead}            & N/A                 & N/A              & N/A                & N/A            & $0.0322$             & $0.0319$          \\ \hline
\textbf{MSE  - Prediction 5 steps ahead}            & N/A                 & N/A              & N/A                & N/A            & $0.0647$             & $0.0627$          \\ \hline
\textbf{RMSE - Prediction 1 step ahead}             & N/A                 & N/A              & N/A                & N/A            & $0.0839$             & $0.0887$          \\ \hline
\textbf{RMSE - Prediction 3 steps aheade}           & N/A                 & N/A              & N/A                & N/A            & $0.1794$             & $0.1785$          \\ \hline
\textbf{RMSE - Prediction 5 steps ahead}            & N/A                 & N/A              & N/A                & N/A            & $0.2544$             & $0.2505$          \\ \hline
\textbf{Average anticipation time of miss deadline} & N/A                 & N/A              & $30,61$            & N/A            & $36,62$              & N/A               \\ \hline
\textbf{Median anticipation time of miss deadline}      & N/A                 & N/A              & $9,44$             & N/A            & $22,60$              & N/A               \\ \hline
\end{tabular}
\end{table}

\begin{figure}[h]
	\centering 
	\subfloat[fig:C3_BoxPlot_TBMD][Anticipation]{\includegraphics[width=6cm]{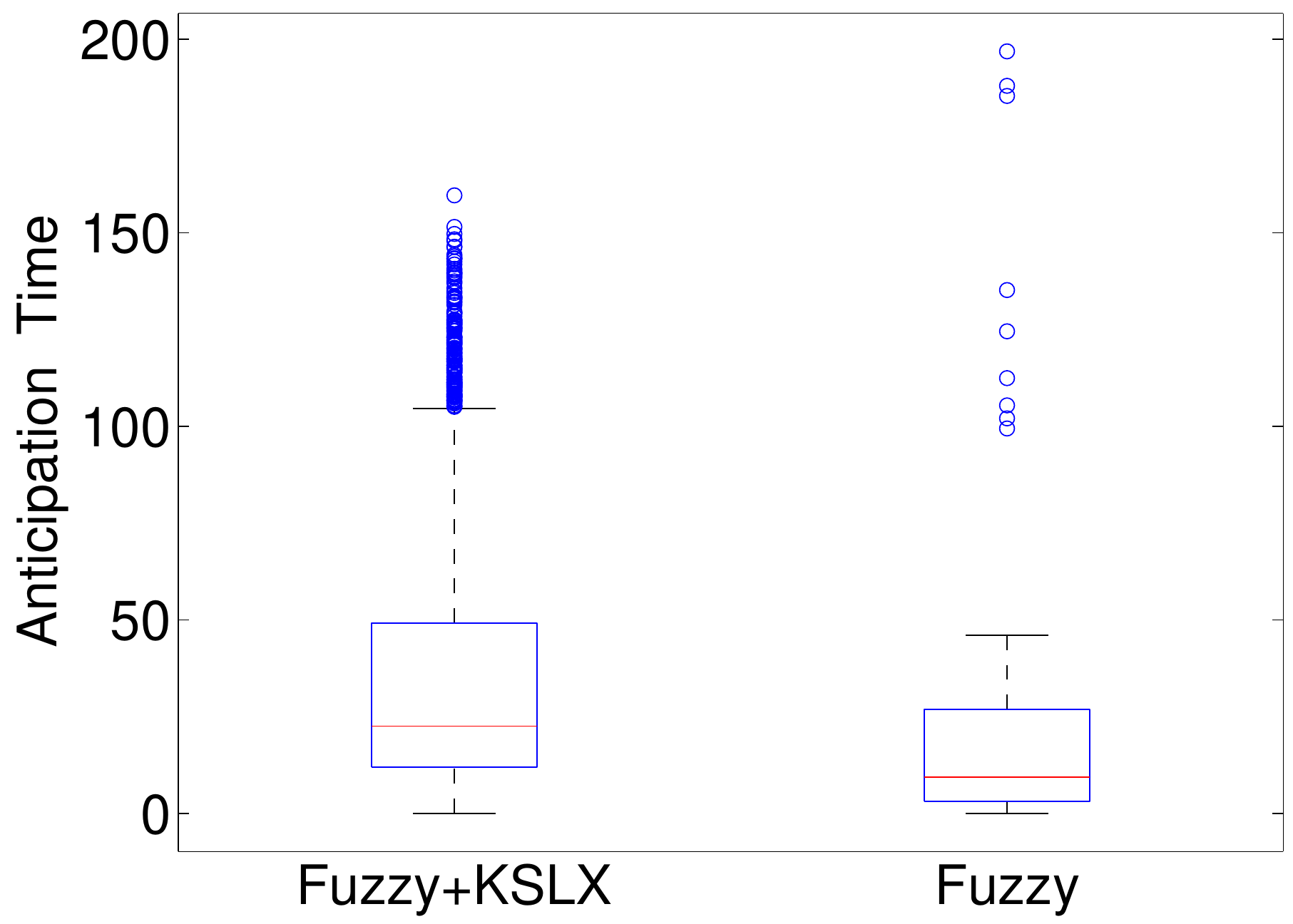}} 
    \hfill
	\subfloat[fig:C3PieMC][Comparison]{\includegraphics[width=6cm]{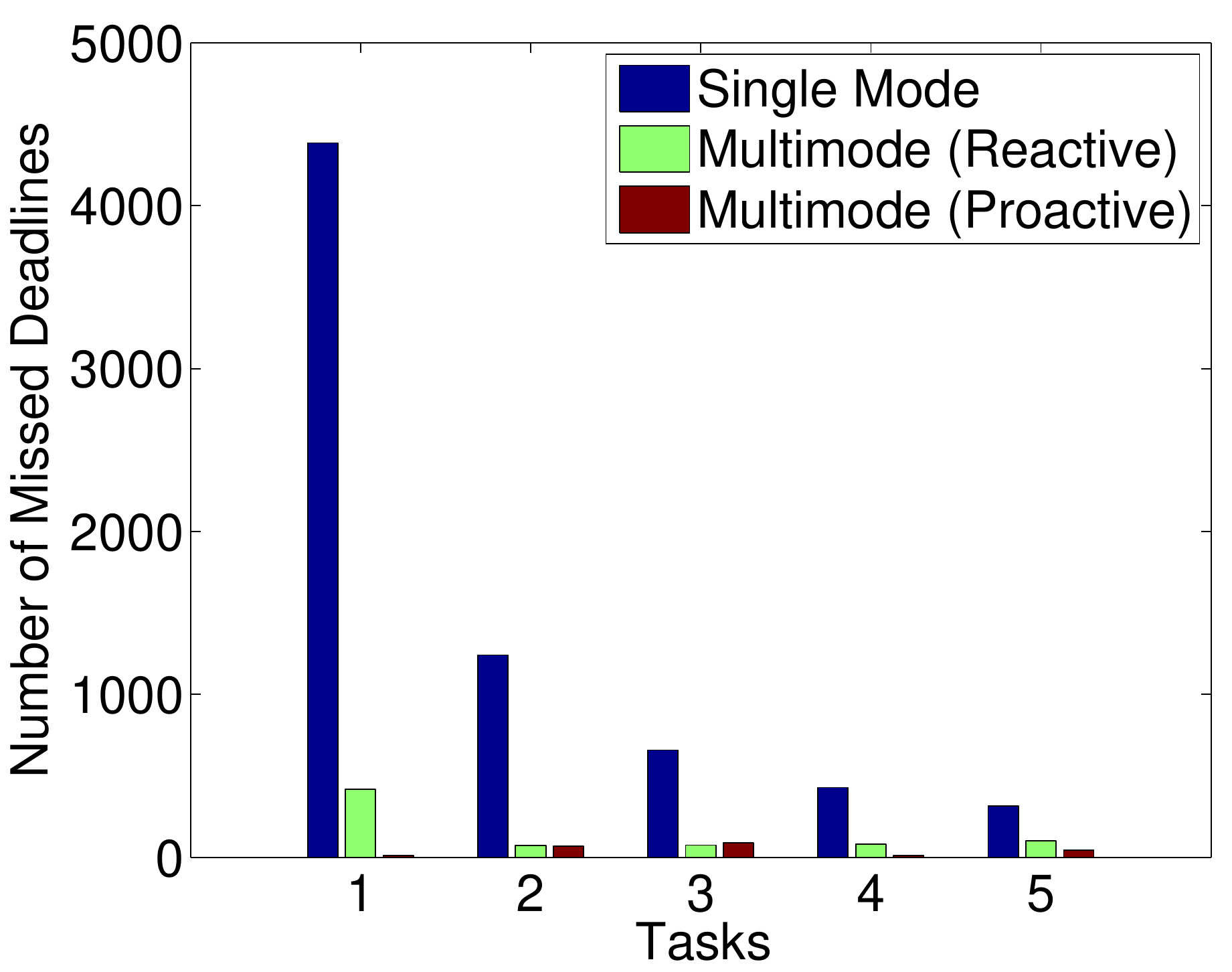}} 
    \caption{\label{fig:C3Chart_BoxPlot_TBMD_LostTasks}Case Study 1 -  Anticipation Time Before Missing a Deadline and Comparison of Missed Deadlines}
\end{figure}

Fig. \ref{fig:C3Chart_BoxPlot_TBMD_LostTasks}a shows the calculated values for the anticipation time before the miss of deadline, with use and without the use of prediction. This time measures the interval between the prediction of miss a deadline and its respective misses within the measurement window.

\begin{figure}[h]
  \centering
  \subfloat[fig:C3FChart1Sept][1 Step Ahead]{\includegraphics[width=6cm]{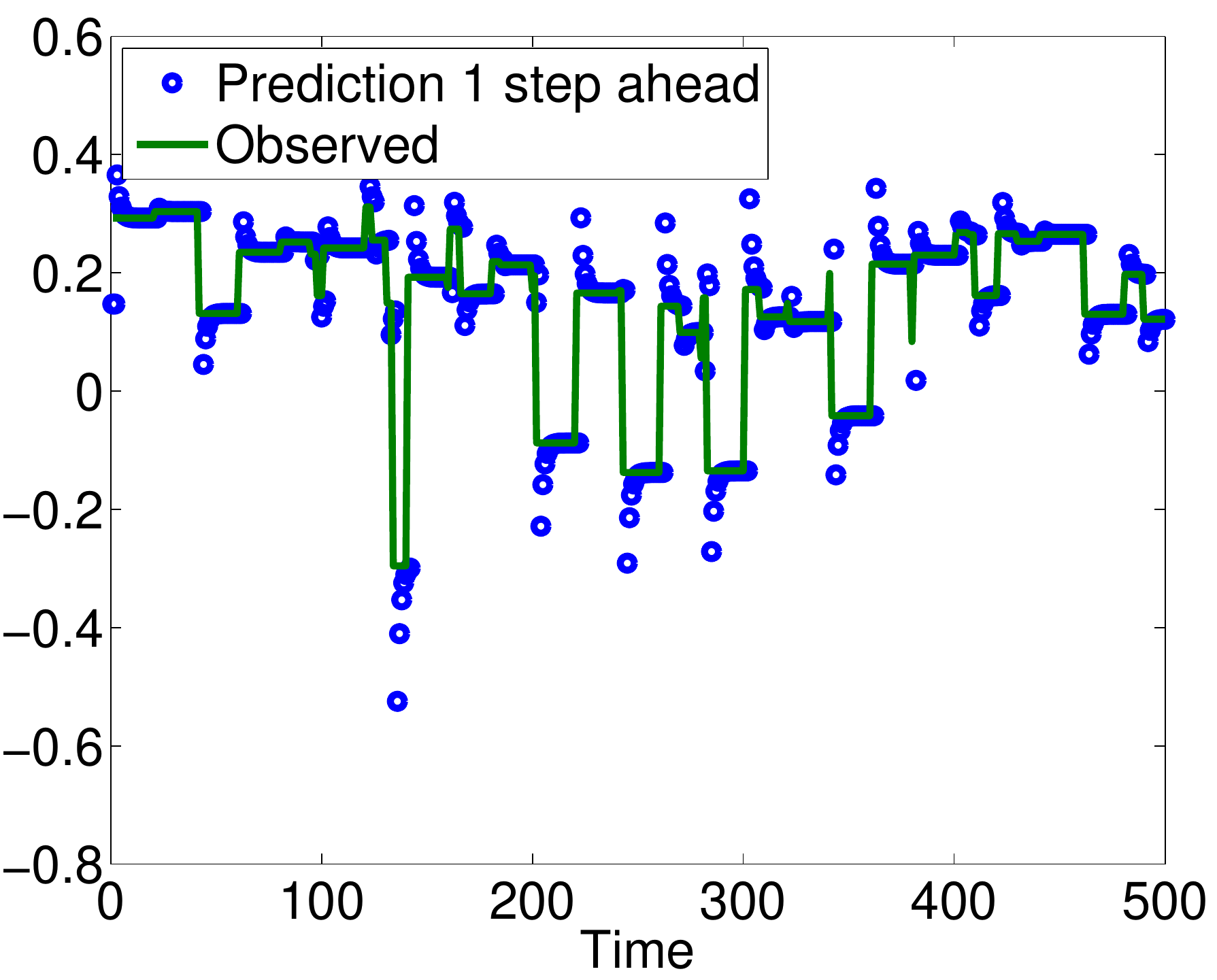}}  
  \hfill
  \subfloat[fig:C3FChart5Sept][5 Steps Ahead]{\includegraphics[width=6cm]{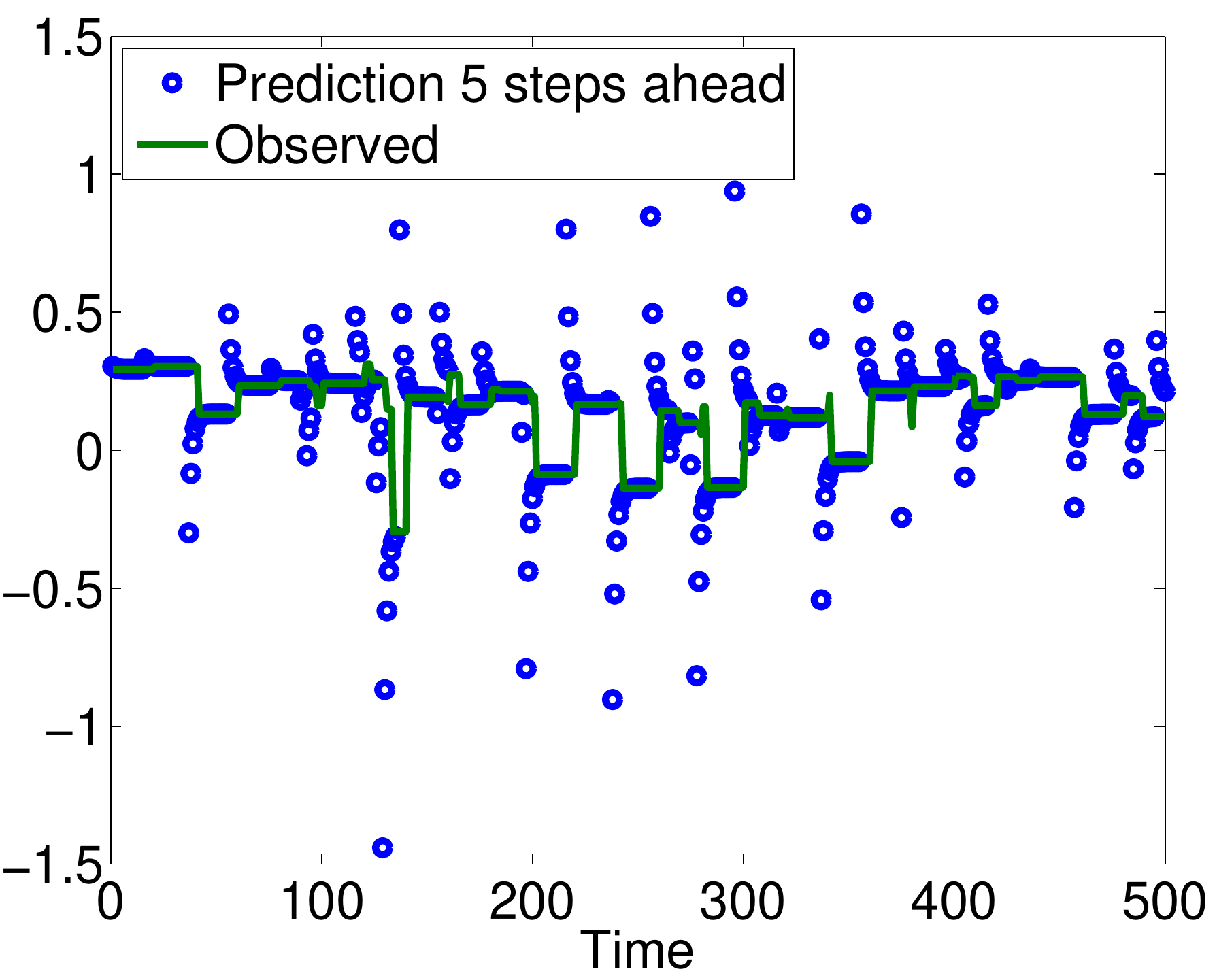}} 
  \caption{\label{fig:C3Chart} Case Study 1 -  Predicted Vs. Observed Value} 
\end{figure}

Figures \ref{fig:C3Chart}a and 
\ref{fig:C3Chart}b present a comparative analysis of the observed values and the predicted values for 1 and 5 steps ahead. Due to the large number of observations the graphs are limited to 500 observations.

Fig. \ref{fig:C3Chart_BoxPlot_TBMD_LostTasks}b presents a comparison, between single-mode and multimode (proactive and reactive) simulations of the number of instances that missed their deadline by task.

\subsection{Case Study 2 - Predicting Mode Changes in FP}
\label{case2}

This case study addresses the same set of tasks described in Section \ref{case1}. However, in relation to the case study presented previously, there was a change in the scheduling policy. In the previous case study, the adopted policy was EDF Preemptive, with Preemptive Fixed Priority being the policy adopted for this case study.

\textit{Assumptions:} The assumptions adopted in this case study were basically the same as in the previous case study described in Section \ref{case1}, with the exception of premise 1, whose scheduling policy was FP.

\textit{Configuration:} As well as the assumptions, also the configurations adopted in this case study are the same as those of the previous case study described in Section \ref{case1}.

\textit{Results:} Table \ref{tab:C4Results} presents the results chosen from the simulation process. In this table, the lines represent the measured variables and the columns present the results of each of the simulation process types executed, namely: 1) reactive mode-change, 2) proactive mode-change, using fuzzy in isolation and 3) change of pro-active critique, combining fuzzy with the predictor KSLX. For each type of simulation process two simulations were performed, the first without changing the criticality level and the second with change.

\begin{table}[h]
\centering
\setlength{\tabcolsep}{0.5pt}
\caption{Case Study 2 - Results}	
\label{tab:C4Results}


\begin{tabular}{|c|c|c|c|c|c|c|}
\hline
\multirow{3}{*}{\textbf{Variables}}                 & \multicolumn{6}{c|}{\textbf{Results}}                                                                                   \\ \cline{2-7} 
                                                    & \multicolumn{2}{c|}{\textbf{Reactive}} & \multicolumn{2}{c|}{\textbf{Fuzzy}} & \multicolumn{2}{c|}{\textbf{Fuzzy+KSLX}} \\ \cline{2-7} 
                                                    & \textbf{Mono}      & \textbf{Multi}      & \textbf{Mono}     & \textbf{Multi}    & \textbf{Mono}       & \textbf{Multi}       \\      \hline
\textbf{CPU utilization rate}                       & $99.89\%$           & $81.72\%$        & $99.89\%$          & $72.60\%$      & $99.89\%$            & $71.88\%$         \\ \hline
\textbf{CPU Busy}                                   & $69.13\%$           & $61.73\%$        & $69.13\%$          & $58.53\%$      & $69.13\%$            & $58.42\%$         \\ \hline
\textbf{CPU Downtime}                               & $30.76\%$           & $19.99\%$        & $30.76\%$          & $14.07\%$      & $30.76\%$            & $13.46\%$         \\ \hline
\textbf{Number of Completed Tasks}                  & $104,050$           & $94,169$         & $104,050$          & $91,171$       & $104,050$            & $91,081$          \\ \hline
\textbf{Number of misses deadlines}                 & $9,758$             & $3,792$          & $9,758$            & $2,298$        & $9,758$              & $1,914$           \\ \hline
\textbf{Percentage  of misses deadlines}            & $9.38\%$            & \$4.03\%'        & $9.38\%$           & $2.52\%$       & $9.38\%$             & $2.10\%$          \\ \hline
\textbf{MSE  - Prediction 1 step ahead}             & N/A                 & N/A              & N/A                & N/A            & $0,0438$             & $0.0286$          \\ \hline
\textbf{MSE  - Prediction 3 steps ahead}            & N/A                 & N/A              & N/A                & N/A            & $5.6209$             & $0.1157$          \\ \hline
\textbf{MSE  - Prediction 5 steps ahead}            & N/A                 & N/A              & N/A                & N/A            & $9.4147$             & $0.2284$          \\ \hline
\textbf{RMSE - Prediction 1 step ahead}             & N/A                 & N/A              & N/A                & N/A            & $0.2093$             & $0.1691$          \\ \hline
\textbf{RMSE - Prediction 3 steps aheade}           & N/A                 & N/A              & N/A                & N/A            & $2.3708$             & $0.3402$          \\ \hline
\textbf{RMSE - Prediction 5 steps ahead}            & N/A                 & N/A              & N/A                & N/A            & $3.0683$             & $0.4780$          \\ \hline
\textbf{Average anticipation time of miss deadline} & N/A                 & N/A              & $146.57$           & N/A            & $146.55$             & N/A               \\ \hline
\textbf{Median anticipation time of miss deadline}  & N/A                 & N/A              & $158.13$           & N/A            & $158.12$             & N/A               \\ \hline
\end{tabular}
\end{table}

\begin{figure}[H]
	\centering 
	\subfloat[fig:C4_BoxPlot_TBMD][Anticipation]{\includegraphics[width=6cm]{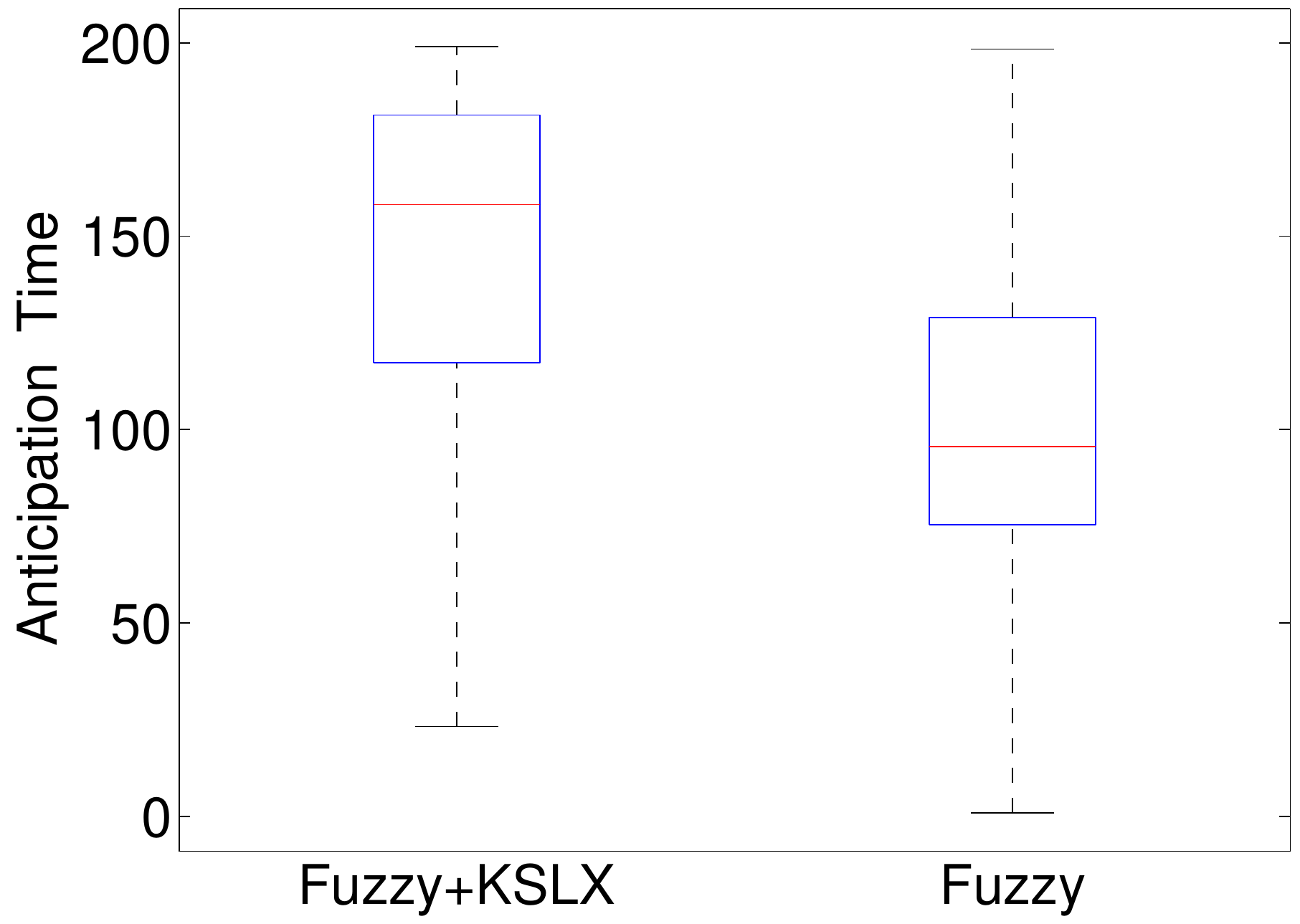}} 
    \hfill
	\subfloat[fig:C4PieMC][Comparison]{\includegraphics[width=6cm]{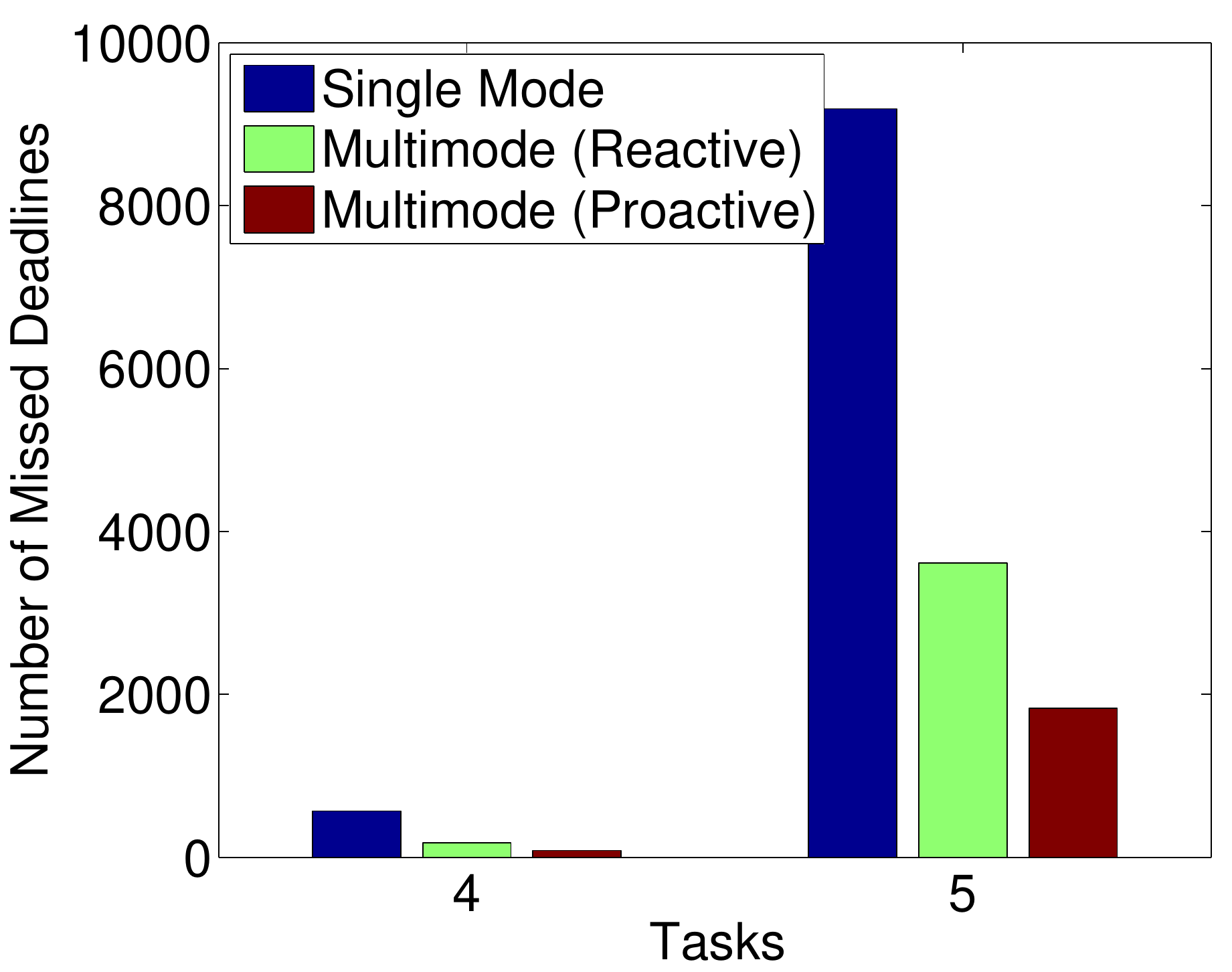}} 
    \caption{\label{fig:C4Chart_BoxPlot_TBMD_LostTasks}Case Study 2 -  Anticipation Time Before Missing a Deadline and Comparison of Missed Deadlines}
\end{figure}

Fig. \ref{fig:C4Chart_BoxPlot_TBMD_LostTasks}a shows the calculated values for the anticipation time before the miss of deadline, with use and without the use of prediction. This time measures the interval between the prediction of miss of deadline and its respective miss within the measurement window.

\begin{figure}[h]
  \centering
  
\subfloat[fig:C4FChart1Sept][1 Step Ahead]{\includegraphics[width=6cm]{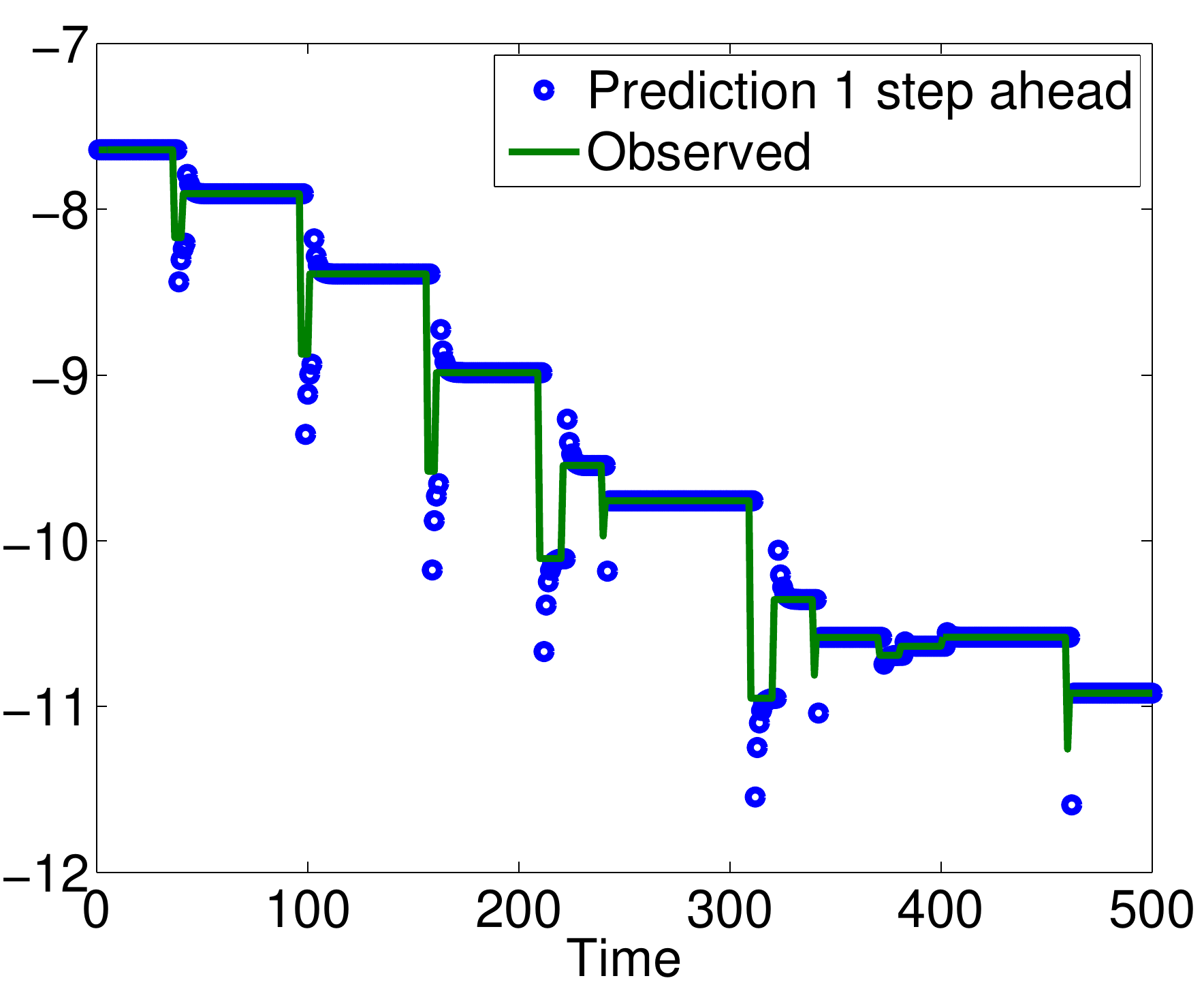}}  
\hfill
\subfloat[fig:C4FChart5Sept][5 Steps Ahead]{\includegraphics[width=6cm]{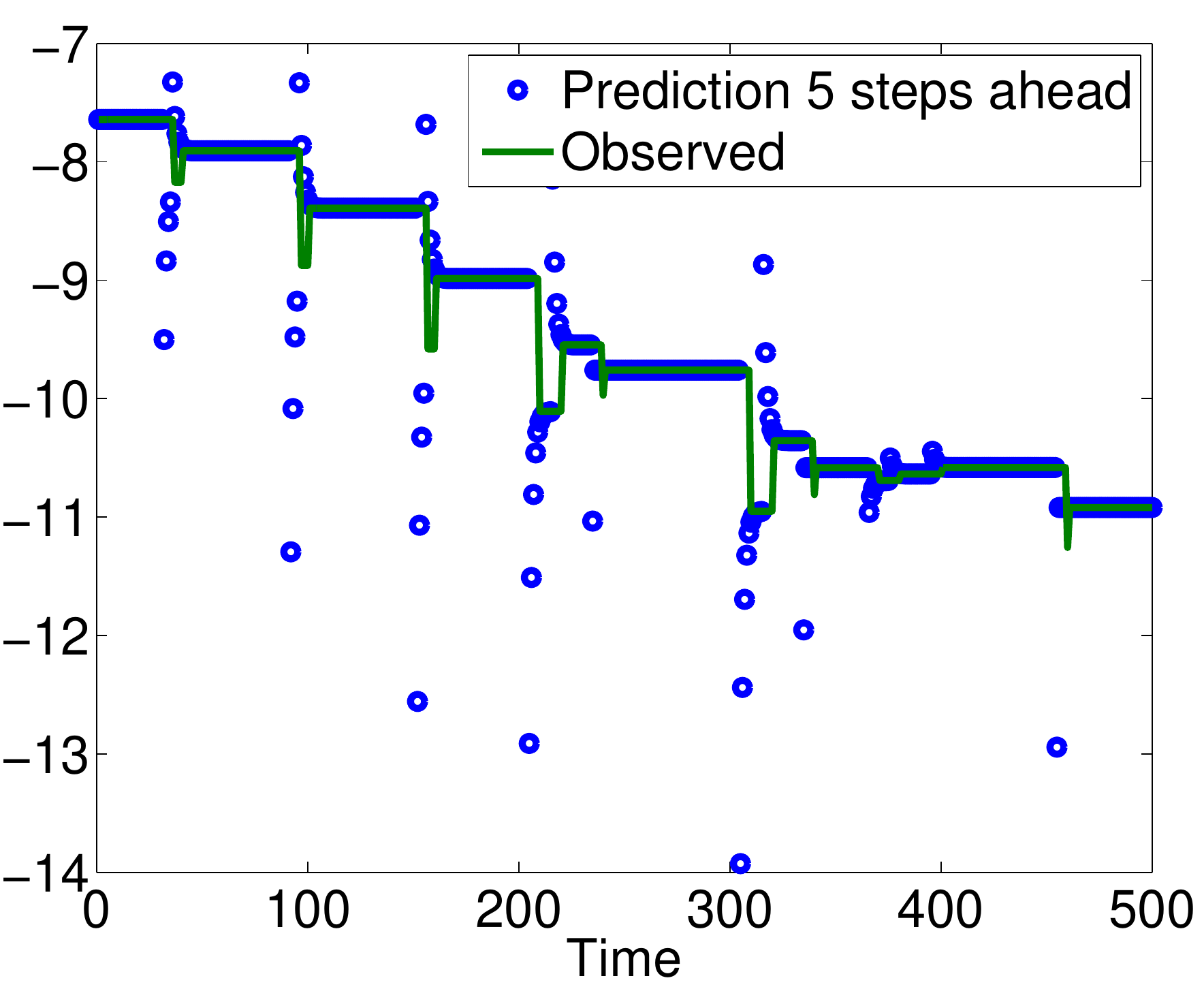}} 
  \caption{\label{fig:C4Chart}Case Study 2 - Predicted Vs. Observed Value}
  
\end{figure}

The Figures \ref{fig:C4Chart}a and \ref{fig:C4Chart}b, present a comparative analysis of the observed values and the predicted values for 1 and 5 steps ahead. Due to the large number of observations the graphs are limited to 500 observations.

Fig. \ref{fig:C4Chart_BoxPlot_TBMD_LostTasks}b presents a comparison, between single-mode and multimode (proactive and reactive) simulations of the number of instances that missed their deadline by task.

\section{Discussion} 
\label{sec:disc}

We now may answer, through the case studies,
the research questions addressed in this work:
1) is it possible to predict the miss of a deadline?
The case studies presented above showed that it is possible to predict the miss of a deadline in
all scenarios studied: Proactive mode changes resulted in better results compared to reactive changes.
2) which candidate variables should be predicted?
The candidate variables for use in prediction were: laxity of each task, the worst case laxity of all tasks and the size of the queue size; however, the variable chosen was the worst case laxity of all tasks, because it was the variable (after an analysis of the generated time series) demonstrated the least mean square error, as can be seen in figures \ref{fig:C3Chart} and \ref{fig:C4Chart},
3) anticipating a mode change can reduce the number of missed deadlines?
it is possible to observe in the case studies that  a significant reduction in the number of missed deadlines occurred when the mode changes were anticipated, i.e. the reduction of missed deadlines was $69.9\%$  in case 1 and $49.5\%$ in case 2 in comparison with reactive mode-changes, 
4) the use of fuzzy in conjunction with the predictor does have better results than using fuzzy in isolation?
the use of fuzzy combined with KSLX predictor provide best results when compared to the use of fuzzy in isolation.  This can be  asserted 
due to the reduction of $52.1\%$ in the number of the missed deadlines in case 1 and $16.7\%$ in case 2,
5) is it feasible to use prediction to identify the miss of deadlines?
the use of prediction to identify the miss of deadlines is  feasible because it allows the anticipation for the need of a mode-change.

In summary, we gathered the following guidelines from this work:

\begin{enumerate}[leftmargin=*]
\item Pro-actively scheduling mode changes is advantageous over reactive scheduling;
\item Reactive scheduling is advantageous over not changing the mode at all;
\item EDF has shown higher performance over fixed-priority when the system includes prediction-based scheduling;
\item The metric used for performance evaluation (number of missed deadlines) and the metric used for scheduling a mode change (worst-case laxity) showed to be  adequate.

\end{enumerate}

\section{Conclusion}
\label{sec:conc}
 
From this work, it can be concluded that the proposed method, which uses prediction in conjunction with fuzzy, is a feasible alternative to avoid  deadline misses in real-time systems with mixed criticality. Proactive scheduling  avoids the need for abruptly aborting tasks; it may allow a delayed or phased abort that 
permit the task to run at least to a safe point. This may be an issue in conventional MCS's where aborts are immediate and tasks do not release resources back to the system.

\bibliographystyle{IEEEtran}

\bibliography{tese}

\end{document}